\documentclass[sigconf, nonacm]{acmart}

\usepackage{url}
\usepackage{ulem}
\usepackage{arydshln} 
\usepackage{subfig}
\usepackage{booktabs} 
\usepackage{amsmath}
\usepackage{amsthm}
\usepackage{amsfonts}       
\usepackage{nicefrac}       
\usepackage{xspace}
\usepackage{xcolor}
\usepackage{backnaur}
\usepackage{multirow}
\usepackage{makecell}
\usepackage{appendix}
\usepackage{enumitem}
\usepackage{balance}
\usepackage{threeparttable}
\usepackage{comment}
\usepackage{amsthm}

\theoremstyle{definition}

\newcommand{\reffig}[1]{Figure~\ref{#1}}
\newcommand{\refsec}[1]{\S\ref{#1}} 
\newcommand{\reftab}[1]{Table~\ref{#1}}

\newcolumntype{R}[1]{>{\raggedleft\let\newline\\\arraybackslash\hspace{0pt}}m{#1}}
\newcolumntype{L}[1]{>{\raggedright\let\newline\\
\arraybackslash\hspace{0pt}}m{#1}}
\newcolumntype{C}[1]{>{\centering\let\newline\\
\arraybackslash\hspace{0pt}}m{#1}}
\def\eg{\textit{e.g.}\xspace}

\def\etc{\textit{etc.}\xspace}
\def\ie{\textit{i.e.}\xspace}

\newtheorem{myDef}{Definition}

\newcommand\vldbpagestyle{plain} 

\begin{document}
\title{ASTA: Learning Analytical Semantics over Tables for~Intelligent~Data~Analysis~and~Visualization}

\author{Lingbo Li}
\authornote{These authors contributed equally to this work.}
\authornote{Work done during an internship at Microsoft Research Asia.}
\affiliation{%
  \institution{Beijing Normal University}
  \institution{ETH Zurich}
  \city{Beijing}
  \country{China}
  \postcode{100875}
}
\email{lingboli@mail.bnu.edu.cn}

\author{Tianle Li}
\authornotemark[1]
\authornotemark[2]
\affiliation{%
  \institution{Hong Kong University of Science and Technology}
  \city{Hong Kong}
  \country{China}
}
\email{tliax@connect.ust.hk}

\author{Xinyi He}
\authornotemark[2]
\affiliation{%
  \institution{Xi'an Jiaotong University}
  \city{Xi'an}
  \country{China}
}
\email{hxyhxy@stu.xjtu.edu.cn}

\author{Mengyu Zhou}
\authornote{Corresponding author.}
\affiliation{%
  \institution{Microsoft Research}
  \city{Beijing}
  \country{China}
}
\email{mezho@microsoft.com}

\author{Shi Han}
\affiliation{%
  \institution{Microsoft Research}
  \city{Beijing}
  \country{China}
}
\email{shihan@microsoft.com}

\author{Dongmei Zhang}
\affiliation{%
  \institution{Microsoft Research}
  \city{Beijing}
  \country{China}
}
\email{dongmeiz@microsoft.com}

\begin{abstract}
Intelligent analysis and visualization of tables use techniques to automatically recommend useful knowledge from data, thus freeing users from tedious multi-dimension data mining. 
While many studies have succeeded in automating recommendations through rules or machine learning, it is difficult to generalize expert knowledge and provide explainable recommendations.
In this paper, we present the recommendation of conditional formatting for the first time, together with chart recommendation, to exemplify intelligent table analysis. 
We propose \textit{analytical semantics} over tables to uncover common analysis pattern behind user-created analyses.
Here, we design analytical semantics by separating \textit{data focus} from \textit{user intent}, which extract the user motivation from data and human perspective respectively.
Furthermore, the \textit{ASTA framework} is designed by us to apply analytical semantics to multiple automated recommendations.
ASTA framework extracts data features by designing signatures based on expert knowledge, and enables data referencing at field- (chart) or cell-level (conditional formatting) with pre-trained models.
Experiments show that our framework achieves recall at top 1 of 62.86\% on public chart corpora, outperforming the best baseline about 14\%, and achieves 72.31\% on the collected corpus ConFormT, validating that ASTA framework is effective in providing accurate and explainable recommendations.

\end{abstract}

\maketitle

\pagestyle{\vldbpagestyle}



\section{Introduction}
\label{sec:intro}

\begin{figure*}
\centering
\includegraphics[width=\linewidth]{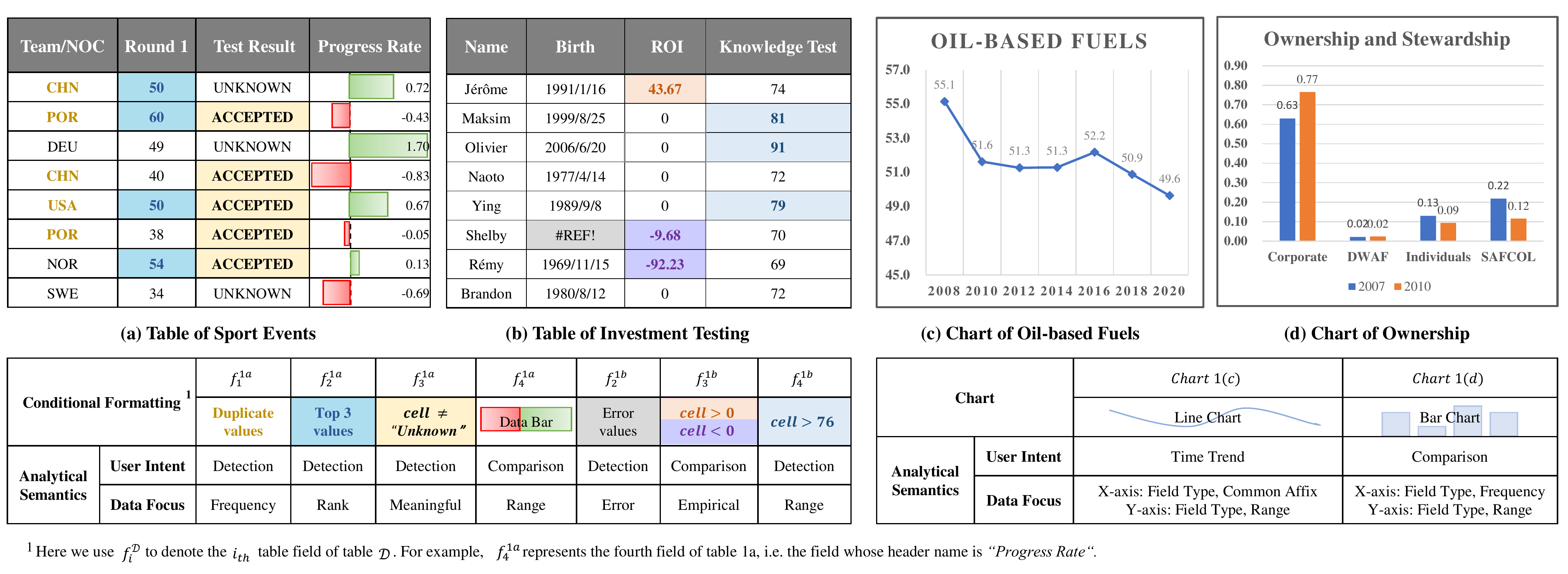}
\vspace{-8mm}
\caption{Examples of conditional formatting and chart together with their corresponding analytical semantics.}
\label{fig:cf_usage}
\vspace{-4mm}
\end{figure*}

Intelligent data analysis and visualization refers to the use of techniques to extract and present useful knowledge from data~\cite{berthold2003intelligent}. Its application on multi-dimensional table becomes popular as it frees people from the painful and tedious process of understanding data. In Excel~\cite{berk2007data}, Google Sheets~\cite{oualline2018using} and Python (\eg, Pandas~\cite{mckinney-proc-scipy-2010} and Matplotlib~\cite{Hunter:2007:Matplotlib}), users create various types of analysis artifacts including conditional formatting, chart, pivot table, formula, \etc, most of which require skill, time and domain expertise to create. 

In this work, we explore the automatic recommendation of conditional formatting and charts to exemplify intelligent table analysis and visualization. Compared to the rich studies on chart creation~\cite{zhou2021table2charts,hu2019vizml,luo2018deepeye,dibia2019data2vis}, the recommendation of \textbf{conditional formatting} is presented for the first time in our study (\refsec{sec:corpus}). 
Conditional formatting enables users to apply special formatting to cells in the spreadsheet that meet certain condition, with examples in \reffig{fig:cf_usage}.
It is a first-class feature in Excel, Google Sheets and Pandas, and there are large amount of public tables with conditional formatted columns created by experts on the web~\cite{abramovich2004spreadsheet,sugden2015conditional,miller2011spreadsheet}. However, conditional formatting has not received much attention.

In the history of research, there has been long line of works~\cite{Zhou:2020:Table2Analysis,garcia2017intelligent, vartak2017towards,Tableau, PowerBI,siddiqui2016effortless,yan2020autosuggest,zhou:2021:table2charts, moritz2018formalizing, hu2019viznet} done to automate table understanding and visual recommendation.
In general, such automation usually involve two steps: Selecting data worthy of being noticed for analysis ({data referencing}), and choosing effective data transformations and presentations ({analysis operation})~\cite{hu2019vizml,zhou:2021:table2charts}.
Thus, the recommendation of conditional formatting and chart can be formulated as operation selection task, data referencing task and complete recommendation task.
This inspired us to design methods that can be applied to recommend multiple analyses.

Unfortunately, existing studies have \textbf{limitations} on following aspects.
1) Generalizing expert knowledge: 
Rule-based systems learn static expert knowledge related to data referencing and operation selection, which is difficult to generalize to other applications~\cite{Wongsuphasawat2016VoyagerEA,viga-lite2016,Vartak2015SeeDBED}.
2) Explaining recommendation results: Though black-box end-to-end machine learning systems provide good analysis suggestions, user often do not know why the generated results are plausible~\cite{hu2019vizml,dibia2019data2vis}.
These limitations not only impede better recommendation, but also hinder us from generating insights into the deep causes behind user-created analyses~\cite{zhang2020explainable,he2015trirank}.

In this view, we propose to explore the motivation behind table analyses and visualizations in terms of \textbf{user intent} and \textbf{data focus},
denoted by us as \textbf{analytical semantics}, which in turn facilitates automatic recommendations. Analytical semantics aims to provide helpful expert knowledge for discovering common analysis patterns, \ie, the \textit{(table / field\footnote{A table field is a logical column of a table as demonstrated in \reffig{fig:cf_usage}.} / cell $\to$ analysis action\footnote{\textit{Analysis action} denotes the actions in table analysis and visualization, including operation selection, data referencing, etc.})} mapping pattern.
In particular, in contrast to existing studies with roughly designed user intent~\cite{mulleveltypo2013,provenance2008,Tessera2021}, we explicitly separate data focus from user intent. Here, {user intent} suggests what purpose motivates people to create analyses over tabular data, while {data focus} investigates which data features are salient and contribute to data referencing. 
Our design makes it possible to explain user behaviors from the perspective of both human being and tabular data. 

Hence, analytical semantics, \ie, the combination of user intent and data focus, provides clear guidance for selecting analysis operations and referencing tabular data, and thus provide recommendations that are better understood by humans.
For example, in chart creation shown in \reffig{fig:cf_usage}c, people use \uline{line chart} (operation selection) to \uline{show time-evolving trends} (user intent), while choosing the \uline{date field with only distinct values} (data focus) as the \uline{x-axis} (data referencing). Another example is conditional formatting.
On the field \textit{"Round 1"} in \reffig{fig:cf_usage}a, one highlights the \uline{top 3 records} (data focus) for \uline{selecting/filtering the podium finishes} (user intent). This pattern \textit{(numerical records of competition results $\to$ select $\textit{top 3}$ cells)} occurs repeatedly in similar table fields, which exemplifies how analytical semantics assists in mining common \textit{(table field $\to$ analysis action)} patterns.

We formulate the automatic recommendation of conditional formatting and chart based on analytical semantics as a machine learning problem, whose inputs are raw tabular data and outputs are analysis actions including user intent, data focus, analysis operation and referenced data in this study.
In order to achieve this process, we identify three challenging problems: First, how to uncover the user intent and data focus behind user-created analysis? Second, how to extract data features based on designed analytical semantics to characterize raw tabular data? Third, how to recommend different types of analyses and visualizations (\eg conditional formatting and chart) via analytical semantics? 

To address the above problems, \textbf{ASTA framework} is proposed by us to learn the \textbf{A}nalytical \textbf{S}emantics over \textbf{TA}bular data and recommend multiple analyses and visualizations. 
Specifically, the ASTA framework is designed as follows:
1) \textit{Analytical semantics design.} In \refsec{sec:analytical_semantics}, we design user intent and data focus based on expert knowledge and corpus investigation to uncover implicit user motivation.
2) \textit{Input feature extraction.} In \refsec{sec:model_input}, we characterize tabular data 
through statistical and linguistic modules: a) Statistic module captures distributional information by extracting metadata features and data signatures designed for mining analytical semantics. The designed signatures also provide a reasonable cell sampling strategy without additional queries or statements.
b) Linguistic module captures semantic information by embedding raw values and contexts through pre-trained models, \eg, RoBERTa~\cite{liu2019roberta} and TABBIE~\cite{iida-etal-2021-tabbie}. 
3) \textit{Multi-task learning by ML model.} In \refsec{sec:ml_model}, we design a machine learning model for the multi-task recommendations, where the analytical semantics learning task assists analysis recommendation tasks.
Compared to the incompetence of previous analysis recommendation studies in cell referencing, ASTA framework enables \textbf{multi-level data referencing} with pre-trained tabular and language models as the backbone, \ie, conditional formatting (cell-level) and chart (field-level) recommendations in this work.

We evaluate ASTA framework on our collected conditional formatting corpus ConFormT and public chart corpora Excel and Plotly (\refsec{sec:experiments}). Results show that our approach achieves high recall numbers on the new problem of conditional formatting recommendation (72.31\% for top 1, and 85.25\% for top 3), and outperforms baselines about 14\% on chart recommendation with a 54.41\% and 62.86\% recall at top 1 for corpus Excel and Plotly respectively.
This demonstrates the applicability and effectiveness of ASTA framework on recommending various analyses and visualizations.
In addition, through ablation study and human evaluations, we show that our model successfully learns analytical semantics over tabular data that are verified to be reasonable by manual annotation.

In summary, our major contributions are:

\noindent $\bullet$ We present a new problem of conditional formatting recommendation and propose analytical semantics by separating data focus from user intent to explore deep causes behind user-created analyses and to facilitate analysis recommendation.

\noindent $\bullet$ We design the ASTA framework to learn analytical semantics and automatically recommend various analyses and visualizations, \eg conditional formatting and chart in this work. It extracts data features through statistical and linguistic modules, and allows multi-level data referencing (cell/field-level) through pre-trained models.

\noindent $\bullet$ We collect a large corpus ConFormT of 289k conditional formatting records from 54k tables. The good performance of ASTA framework is evaluated on ConFormT and public chart corpora.

\section{Conditional Formatting}
\label{sec:corpus}

For data mining and information sharing, people create a large amount of analyses and visualizations over tables, \eg, conditional formatting, chart and pivot tables, in Excel spreadsheets and Google sheets. 
We have chosen charts and conditional formatting as typical examples.
Since charts are the most common practise today for table visualization and have spawned many related studies, we briefly introduce charts and show how our method can be adapted to chart recommendation in \refsec{sec:chart}. In contrast, the problem of conditional formatting recommendation is presented for the first time in this paper. We introduce conditional formatting and a new large corpus ConFormT we collected as follows.



\textbf{Conditional formatting} is an analysis artifact on tabular data, which enables users to provide customized conditions to select data and determine how to format them. 
We display many examples in \reffig{fig:cf_usage} to demonstrate the conditional formatting applied to table fields, for instance, \textit{"highlight error cells in bold font and grey fill"} on field $f_2^{1b}$. It can be seen that conditional formatting provides users with a variety of conditions (\eg \textit{"top k values"} and \textit{"is duplicate values"}) and formats (\eg \textit{"font"} and \textit{"fill"}). We have summarized these forms into a few categories shown in Table \ref{tab:conditional_formatting}.

Table \ref{tab:conditional_formatting} shows an illustration of how conditional formatting is organized in commercial systems, as well as our induction. 
The options for conditional formatting are categorized by us into 12 operations (\textit{Operation}), each of which needs to be paired with $0$ or $n$ parameters ($\#P$).
For the sake of clarity, each operation is supplemented with its conditional form and corresponding examples in \reffig{fig:cf_usage}. 
Note that we do not focus on formatting details in this work, which rely heavily on user preferences rather than on data characteristics, \eg, in \reffig{fig:cf_usage}, both $f_2^{1a}$ and $f_3^{1a}$ aim to highlight the selected data, although their formats are different colors.

\begin{table}
  \centering
  \footnotesize
  \begin{threeparttable}
    \begin{tabular}{p{0.15cm}|p{2.95cm}|p{0.4cm}|p{2.85cm}|p{0.45cm}}
    \toprule
     & \textbf{Operation} & \textbf{\#P\tnote{1}} & \textbf{Condition Form\tnote{2}} & \textbf{Ref.}  \\
    \midrule
    1 & Is (Not) Error & 1 & \textit{$v_i =$ / $\neq$ error values} & $f_2^{1b}$     \\
    \midrule
    2 & Is (Not) Blank & 1 &\textit{$v_i =$ / $\neq$ blank value} & -   \\
    \midrule
    3 & Is Duplicate & 0 & \textit{frequency ($v_i$) > 1} & $f_1^{1a}$      \\
    \midrule
    4 & Less / Greater Than (Or Equal) & 1 & \textit{$v_i <$ / $>$ / $\geq$ / $\leq p$} & $f_4^{1b}$    \\
    \midrule
    5 & Top (Bottom) K & 1 & \textit{max-k / min-k values} & $f_3^{1a}$      \\
    \midrule
    \multirow{2}{*}{6}  & \multirow{2}{*}{(Not) Between} &  \multirow{2}{*}{2} & \textit{$v_i >$ / $\geq p_A$ and $v_i <$ / $\leq p_B$} & \multirow{2}{*}{-}  \\
    & & & \textit{$v_i <$ / $\leq p_A$ or $v_i >$ / $\geq p_B$} &  \\
    \midrule
    7 & (Not) Equal / Contains\tnote{3} & 1 & \textit{$v_i =$ / $\neq p$, $p \in$ / $\notin v_i$} & $f_3^{1b}$    \\
    \midrule
    8 & (Not) Equal Set & $\geq 2$ & \textit{$v_i =$ / $\neq p_A$, $v_i =$ / $\neq p_B$, $\cdots$} & -      \\
    \midrule
    9 & Data Bar & \multirow{4}{*}{\makecell[l]{\\ $\geq 2$}} & \multirow{4}{*}{\makecell[l]{\\ \textit{$p_A <$ / $\leq v_i <$ / $\leq p_B$,} \\ \textit{$p_B <$ / $\leq v_i <$ / $\leq p_C$,} \\ \textit{\dots}}} & \multirow{4}{*}{\makecell[l]{\\ $f_4^{1a}$\\ \\ $f_3^{1b}$}}   \\
    \cmidrule(){1-2}
    10 & Color Scale & & & \\
    \cmidrule(){1-2}
    11 & Icon Set & & &  \\
    \cmidrule(){1-2}
    12 & Partition Set & & &  \\
    \bottomrule
    \end{tabular}%
    \begin{tablenotes}  
        \small 
        \item[1] $\#P$ records the number of parameters required to complete an operation.
        \item[2] \textit{Condition Form} records formula expressions of conditions corresponding to the operation. $p$ and \textit{k} are parameters provided by users.
        \item[3] More than 90\% of the parameters of operation \textit{(Not) Contains} are complete cell values, which are equivalent to the operation \textit{(Not) Equal}. Therefore, these two operations are merged into one type in our solution.
    \end{tablenotes} 
  \end{threeparttable}
  \caption{Conditional formatting in commercial systems (\eg Excel and Google Sheets). Action space of conditional formatting consists of \textit{operation} actions and \textit{parameter} actions.   \label{tab:conditional_formatting}}
  \vspace{-7mm}
\end{table}

\subsection{Recommendation Problem} 
\label{sec:cf_actions}

We first define a table and an analysis and visualization.
A \textit{table} here is an $n$-dimensional dataset $T$ which contains $n$ table fields $\mathcal{F}^{T} = (f_1^T,\cdots,f_n^T)$. Each logical column from tables in \reffig{fig:cf_usage}, for instance, is a table field $f$ with its first row as header. An \textit{analysis and visualization} is a combination of $m$ actions from the action space $\mathcal{A} = (a_1,\cdots,a_k)$. Referring to ~\cite{hu2019vizml, dibia2019data2vis, zhou:2021:table2charts}, we regularize the action space of conditional formatting (and chart in \refsec{sec:chart}) as the actions of operation types $\mathcal{O} = \{o_i\}$ and referenced data $\mathcal{R} = \{r_j\}$, i.e. $\mathcal{A} = \mathcal{O} \cup \mathcal{R} = (o_1,\cdots,o_i,r_1,\cdots,r_j)$.
We then give the definition of conditional formatting as follows.
\begin{myDef}
(Conditional Formatting) An executable conditional formatting record on table field $f$ requires both an operation $o$ and an optional parameter set $\{r\}$:
\begin{equation}
\begin{aligned}
CF(f):= \{ o, r_1, \cdots, r_k \mid o \in \mathcal{O}_{CF}, r \in \mathcal{R}_{CF}, \; s.t. \; k \in \mathbb{N} \}
\end{aligned}
\end{equation}
\end{myDef}

\noindent Thus, the recommendation problem of conditional formatting involves operation selection and parameters generation.

\subsubsection*{\textbf{Operation ($\mathcal{O}_{CF}$)}}
Commercial platforms such as Excel and Google Sheets provide users with many conditional formatting operations in a similar way. We focus on the most commonly used operations, which account for more than 95\% cases in the corpus ConFormT. The operation space of conditional formatting $O_{CF}$ in this study - 12 operations in total - is demonstrated in Table \ref{tab:conditional_formatting}. Note that we try to eliminate the subjective factors in this problem by merging operations that have complementary effect. For example, $o_i = "Equal"$ and $o_j = "Not$ $Equal"$ extract the complementary sets of cells $V_i$ and $V_j$ from the given table field, indicating that whether users select $V_i$ or $V_j$ depends heavily on personal preferences. These subjective effects are avoided by merging the two into one category.

\subsubsection*{\textbf{Parameter ($\mathcal{R}_{CF}$)}}
As described in Table \ref{tab:conditional_formatting}, parameter(s) is required to complete a condition form else for the operation \textit{"Is Duplicate"}. For instance, when people use operation \textit{"Greater Than"} (i.e., $v_i \geq p$) to perform conditional formatting, the parameter $p$ should be provided. A parameter $r_i(f) \in \mathcal{R}_{CF}$ can be a number or text according to the operation $o(f)$ and table field $f$, for example, number $76$ for numerical field $f_4^{1b}$. The parameter space $\mathcal{R}_{CF}$ in this study includes cell values of the given table, as well as possibly some customization options as described in \refsec{sec:analytical_semantics}, \eg the average number.

\subsection{Conditional Formatting Corpus} 
\label{sec:cf_corpus}
To evaluate the recommendation tasks of conditional formatting, we construct a large-scale corpus ConFormT (\textit{Conditional Formatting Tables}), which is the first corpus containing diverse tables and conditional formatting records.

\subsubsection*{\textbf{Corpus Preparation.}}
Our corpus ConFormT is extracted from Excel spreadsheet files crawled from the public web. The following steps of data preparation are taken in this study:

1) \textit{Raw record extraction.} 
Extract the raw records of conditional formatting from Excel spreadsheet via OpenXML3~\footnote{ Open XML SDK. See https://github.com/OfficeDev/Open-XML-SDK.}. The records will be merged if they reference the same region and apply the same analysis (\ie, same operation and parameters).

2) \textit{Source table restoration.} 
Extract the source tables of conditional formatting records from Excel spreadsheet via table detection algorithm~\cite{dong2019semantic}. The simple tables are retained while combo tables are discarded. The records will be dropped if its references are not covered by any detected simple table.

3) \textit{Table and record Matching.} 
Split the records, whose references cover more than one table, into several records according to the table range and direction. Then match all of the records with tables based on their references to construct (record, table) pairs.

4) \textit{Record merging and filtering.} 
Merge the records with same operation and parameters that are applied to the same table field, to produce new records and (record, table) pairs. Filter the (record, table) pairs if the proportion of cells referenced by the record in corresponding table field is below the given threshold.

5) \textit{Table deduplication and down sampling.} 
Group tables based on their schemas to avoid the “data leakage” problem that duplicated tables are allocated into both training and testing sets. Randomly sample at most 5 unique tables for each unique (schema, table) pair to mitigate imbalanced schemas.


\vspace{-3mm}
\begin{figure}[!h]
	\centering
	\includegraphics[width=\columnwidth,trim=0 0 0 30,clip]{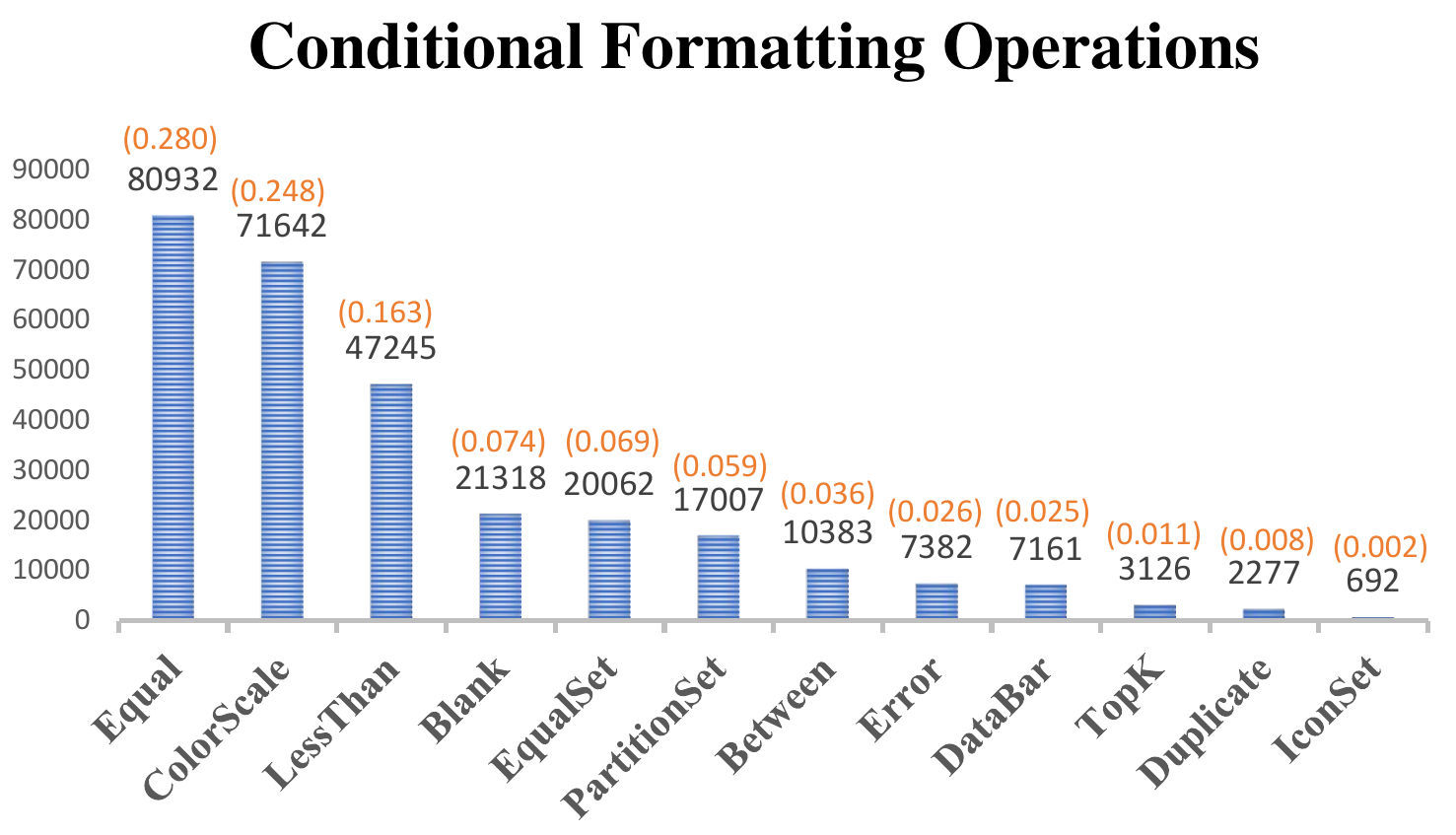}
	\vspace{-7mm}
	\caption{Statistics of conditional formatting operations in the collected corpus ConFormT. 
	\label{fig:corpus_stat}}
	\vspace{-4mm}
\end{figure}

\subsubsection*{\textbf{Corpus Statistics.}}
By the preparation procedure, we collect 54,248 tables and 289,227 conditional formatting records from original Excel files. These table have 150 rows and 21 columns in average. Conditional formatting fields account for 20\% - 40\% of a table in more than half of the cases. As shown in Figure \ref{fig:corpus_stat}, our corpus ConFormT contains 12 operations, including 80932 \textit{Equal}, 71642 \textit{Color Scale}, 47245 \textit{L/G Than}, 21318 \textit{Is Blank}, 17007 \textit{Partition Set}, 20062 \textit{Equal Set}, 10383 \textit{Between}, 7161 \textit{Data Bar}, 7382 \textit{Is Error}, 2277 \textit{Is Duplicate}, 3126 \textit{Top K}, and 692 \textit{Icon Set}. The collected corpus ConFormT is used for model training and evaluation.

\section{Methodology}
\label{sec:method}

We design a novel framework ASTA to learn analysis semantics on tabular data and to address the recommendation of conditional formatting and chart. 

\subsection{ASTA Framework Architecture}

\begin{figure}
\vspace{-2mm}
	\centering
	\includegraphics[width=0.7\columnwidth]{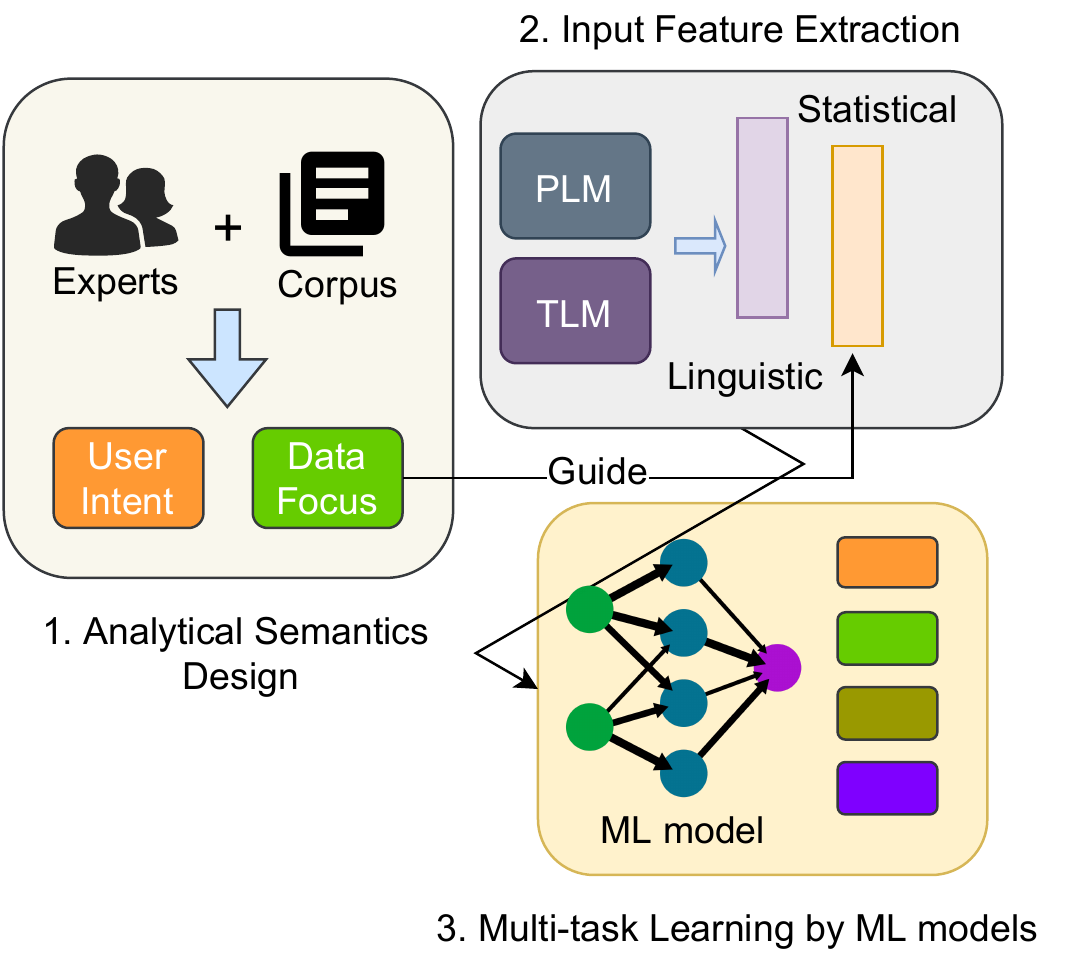}
	\vspace{-3mm}
	\caption{Overall architecture of ASTA framework.
	\label{fig:framework}} 
	\vspace{-3mm}
\end{figure}

We present the ASTA framework, the first framework that aims to resolve the analysis and visualization problems over tabular data based on pre-trained tabular models. The overall architecture of our framework is displayed in Figure \ref{fig:framework}.
First, we design task-specific analytical semantics according to corpus investigation and expert knowledge. Second, based on the data focus, we design distributional signatures to extract statistical features of raw tabular data, and capture linguistic information leveraging pre-trained tabular model and language model. 
Finally, ASTA combines distributional and linguistic information with analytical semantics learning sophisticatedly. We can obtain the predictions of analysis operation and referenced data as the recommendation results. This paradigm allows recommendation on different structure levels in tabular data and performances better than the previous SOTA models. 

\subsection{User Intent and Data Focus Design}
\label{sec:analytical_semantics}

\begin{figure*}
\centering
\includegraphics[width=\linewidth]{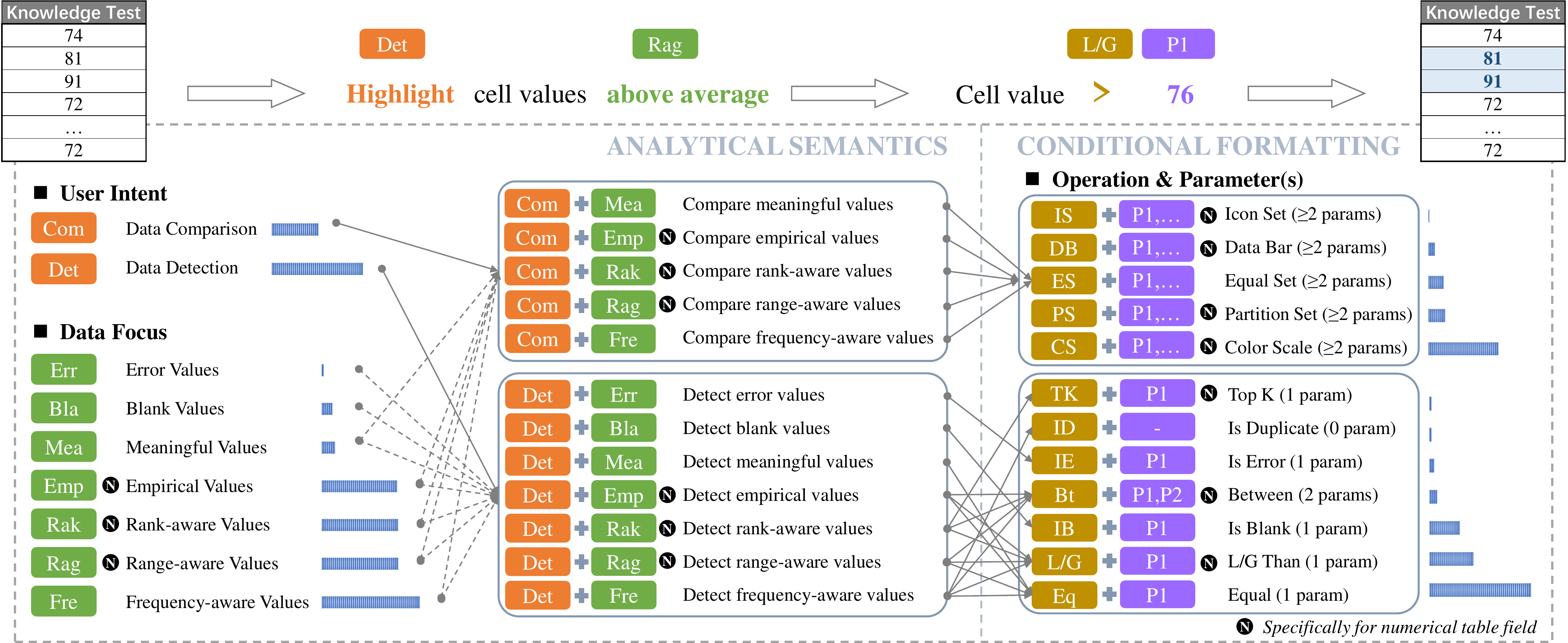}
\vspace{-6mm}
\caption{Analytical semantics for conditional formatting and an illustration of the way it functions. }
\vspace{-3mm}
\label{fig:analytical_semantics}
\end{figure*}

We propose to explore the motivation of users to create table analysis and visualization in terms of both user intent and data focus, denoted by us as \textbf{analytical semantics} in this work.
\begin{myDef}
(Analytical Semantics) Analytical semantics paradigm consists of user intent $\mathcal{U}$ and data focus $\mathcal{D}$:
\begin{equation}
\begin{aligned}
AS:= \{ u_1, \cdots, u_i, d_1, \cdots, d_j \mid u \in \mathcal{U}, r \in \mathcal{D}, s.t. \; i, j \in \mathbb{N}^+ \}\\
\end{aligned}
\end{equation}
\vspace{-5mm}
\end{myDef}
\noindent Here, \textit{user intent} indicates what purpose motivates people to create analyses over tabular data, while \textit{data focus} investigates which data features are salient and contribute to data referencing. Unlike existing studies where user intent was presented in general terms~\cite{mulleveltypo2013,provenance2008,Tessera2021}, we explicitly separate data focus from user intent, making it possible to explain the deep causes behind user-created analysis from the perspective of both human being and tabular data.


In the first step of ASTA framework, we design user intent and data focus to better understand the specific table analysis and visualization, \eg, conditional formatting and chart. As shown in \reffig{fig:analytical_semantics}, we take conditional formatting as an example to show how analytical semantics (\ie, user intent and data focus) functions. See the analytical semantics for chart recommendation in \refsec{sec:chart}.

\subsubsection*{\textbf{User intent ($\mathcal{U}_{CF}$).}}
\label{sec:user_intent}

Referring to previous studies on other visualization tasks\cite{Tessera2021,mulleveltypo2013,provenance2008}, we designed the following user intents for the newly proposed conditional formatting problem, \ie, 
\begin{equation}
\begin{aligned}
\mathcal{U}_{CF} := \{u_i, i \in [1,2] \} := \{ \textit{Det} , \; \textit{Com} \} 
\end{aligned}
\end{equation}

\noindent Here, $\textit{Data}$ $\textit{Detection}$ (\textit{Det}) indicates that people select partial data from a table field and highlight them by self-defined formats, \eg field $f_1^{1a}$, $f_2^{1a}$ and $f_4^{1b}$. By contrast, $\textit{Data}$ $\textit{Comparison}$ (\textit{Com}) refers to the situation that people wanna compare multiple sets of cells by marking them with different formats, \eg field $f_4^{1a}$ and $f_3^{1b}$. 
Figure \ref{fig:analytical_semantics} displays the quantity comparison of these two intents in the corpus ConFormT using the blue bars.

We generate the golden labels for user intent of user-created analyses based on expert-designed heuristic rules involving operation and record number\footnote{If two records of \textit{"Equal"} operation are applied to the same table field, the user intent will be \textit{Data Comparison}, while only one "Equal" record indicates \textit{Data Detection}.}. This requires us only to optimize the rules and thus reduces the manual burden of labelling samples.

\subsubsection*{\textbf{Data focus ($\mathcal{D}_{CF}$).}}
\label{sec:data_focus}
In order to dig out the potential reasons why users exert specific operations on designated referenced data but not others, we define and classify data focus to further reveal the latent motif users are likely to have from the perspective of tabular data features. Specifically, we determine the subset of data focus limited to this work heuristically summarized as the following steps.

1) \textit{Investigating Data.} We invite three experts in domain, each to go through 100 samples randomly selected from ConFormT. 

2) \textit{Listing Potential Candidates.} The experts conclude and list the potential reasons why people make their visualization choice with the specific parameters according to their expert knowledge and data characteristics in terms of numeric distribution and textual semantics, \eg, the top-frequency values and the meaningless values "None". During this step, the experts need to construct concrete principles for each candidate, and determine specific thresholds, \eg, the top-frequency values refer to the cells whose frequency out of the total cell numbers in the target field is larger than 30\%.

3) \textit{Verifying Design Coverage.}
We conduct statistical analysis on each of the candidate, and calculate the coverage in the ConFomT corpus. Finally, we select the top ones in order until they can cover 90\% of the cases in the complete dataset.

The final candidates of $\mathcal{D}_{CF}$ are as follows.
\begin{equation}
\begin{aligned}
\mathcal{D}_{CF} := \{d_{i}, i \in [1,7]\} = \{\textit{Err},\textit{Bla}, \textit{Mea}, \textit{Emp},\textit{Rak}, \textit{Rag}, \textit{Fre} \} 
\end{aligned}
\end{equation}

\noindent As shown in \reffig{fig:analytical_semantics}, we explain the seven data focuses by giving corresponding examples\footnote{Note that $\{ \textit{Rak}, \textit{Rag}, \textit{Emp} \}$ apply only to numerical fields, and the others are available for both string and numerical fields.}:



\noindent $\bullet$  \textit{Frequency (Fre)}: 
In $f_1^{1a}$, the values that occur more than once are shown in yellow, \ie, $\textit{frequency}(v_i) > 1$.

\noindent $\bullet$  \textit{Range-aware (Rag)}: 
In $f_4^{1b}$, the values above average are highlighted in blue, \ie, $v_i>\textit{mean}(\mathcal{V}_{f_4^{1b}})$.

\noindent $\bullet$  \textit{Rank-aware (Rak)}:
In $f_2^{1a}$, the top 3 values after sorting in descending order are highlighted in red fill, \ie, $e_i \in \textit{max-3}(\widehat{\mathcal{V}}_{f_3^{1a}})$.

\noindent $\bullet$  \textit{Meaningful (Mea)}:
If cell values of the given filed $\mathcal{V}_f=$ \textit{\{"English", "Physics", "None"\}}, people often pick out the \textit{"None"} cells and format them as white to hide these records, in the corpus ConFormT. 
In $f_3^{1a}$, \textit{"ACCEPTED"} is distinguished from the meaningless \textit{"Unknown"}.

\noindent $\bullet$  \textit{Error (Err)}:
In $f_2^{1a}$, the error cells (\textit{"\#REF!"}) are emphasized.

\noindent $\bullet$  \textit{Blank (Bla)}:
Users distinguish blank cells with others by formats.

\noindent $\bullet$  \textit{Empirical (Emp)}:
In $f_3^{1b}$, the parameter $0$ is a common splitting rule for "ROI" calculation. Based on expert knowledge, some numbers, \eg, 0 and 1, are used frequently in data segmentation and thus are considered as empirical values in this work. 

It can be seen that these categories reflect diverse data features and can also be used to differentiate data. As shown in \reftab{tab:answer_span}, we collect common signatures (\ie, \textit{Common Frequency, Common Rank} and \textit{Common Range}) and common vocabulary of meaningless values and empirical values for cells based on ConFormT. The common signatures of fields are designed for chart creation in \refsec{sec:chart}.

We produce the golden labels for data focus by comparing the user-created parameters in ConFormT with the collected common signatures and vocabulary.
We compare them to determine which one or multiple data focuses the conditional formatting applied to the target field belongs to. For example, in \reffig{fig:analytical_semantics}, the parameter $76$ for operation \textit{Less/Greater Than} is the average of cell values in field $f_4^{1b}$. Given that the average value ($\bar x$) belongs to the common \textit{Range-aware} values defined in \reftab{tab:answer_span}, the conditional formatting on field $f_4^{1b}$ is classified as data focus \textit{Rag}.

\vspace{-1mm}
\subsubsection*{\textbf{Operation and parameter.}} 
\label{sec:ope_param}

As demonstrated in Figure \ref{fig:analytical_semantics}, the combination of user intent and data focus provides clear guidance for specifying operations and parameters for the target tabular data.
In \reffig{fig:analytical_semantics}, for instance, when people tend to highlight partial values (\textit{Det}) and give priority to the range feature of the target field (\textit{Rag}), \textit{"Detect Range-aware Values"} leads to alternative operations (\ie, "\textit{Equal}", "\textit{Between}" and "\textit{Less/Greater Than}") and several range-aware parameters (\textit{Rag} cells, \eg, the average value 76 and the $(\max x + \max x)/2$ value 80).



\vspace{-1mm}
\subsubsection*{\textbf{Machine Learning Tasks.}} 
\label{sec:ml_formulation}

We formulate our problem as the following machine learning tasks:

1) \textit{Analytical semantics learning.} Given a table $T$ and table field $f$, we generate the top-1 recommendation of user intent $u_i$ by probability $P(u_i \in \mathcal{U}_{CF} \mid T, f)$ and the top-k recommendations of data focus $(d_1,\cdots,d_k)$ by probability $P(d_i \in \mathcal{D}_{CF} \mid T, f)$. 
In order to obtain completed analytical semantics, we also generate a top-k recommendation list $((u_1, d_1),\cdots,(u_k, d_k))$ ranked by probability $P((u_i, d_i), u_i \in \mathcal{U}_{CF}, d_i \in \mathcal{D}_{CF},  \mid T, f)$

2) \textit{Conditional formatting recommendation.} 
Conditional formatting recommendation includes three tasks, \ie, operation classification, reference generation and complete conditional formatting recommendation. 
Given a table $T$, table field $f$ and the learned analytical semantics $(u,d)$, these tasks are formulated as follows.

\noindent $\bullet$ \textit{Operation classification task.} Generate a top-k list of operations $(o_1,\cdots,o_k)$ ranked by probability $P(o_i \in \mathcal{O}_{CF} \mid T, f, u,d)$. 

\noindent $\bullet$ \textit{Reference generation task.} Generate a top-k list of parameters $(r_1,\cdots,r_k)$ ranked by probability $P(r_i \in \mathcal{R}_{CF} \mid T, f, u,d)$. 

\noindent $\bullet$ \textit{Complete analysis recommendation task.} Generate a top-k list of conditional formatting records $\{(o_i,r_i)\}$ ranked by probability $P(o_i \in \mathcal{O}_{CF}, r_i \in \mathcal{R}_{CF} \mid T, f, u,d)$. 


\begin{table}
\footnotesize
  \centering
  \begin{threeparttable}
    \begin{tabular}{C{2.2cm}|L{5.5cm}}
    \toprule
    \textbf{Data Focus\tnote{1}} & {\makecell[c]{\textbf{Common Signatures of Cells}}} \\
    \midrule
    \makecell[c]{{Meaningless\tnote{2}}} & {\makecell[l]{$\{x\}$, where $x \in V^T$ and $x \in \mathcal{M}$}} \\
    \midrule
    \makecell[c]{{Empirical\tnote{3}}}  & {\makecell[l]{$\{x\}$, where $x \in \mathcal{E}$}} \\
    \midrule
    \makecell[c]{{Rank-aware}}  & {\makecell[l]{$x$ at the position 1, 3, 5, 10, 20 and $10 \cdot$ $n (n \in \mathbb{Z}^+,$ $1$ \\ $ \leq n\leq 9)$ percentiles after descending or \\ ascending sorting, where $x \in V$}} \\
    \midrule
    \makecell[c]{{Range-aware}}  & {\makecell[l]{$\bar x$, ($\max x - \min x$) / 2 and $x$ close to the integer \\ multiples, where $x \in V$}} \\
    \midrule
    \makecell[c]{{Frequency-aware}}  & {\makecell[l]{$x$ whose $\textit{freq}(x)$ is in the top $30\%$ after ascending \\or descending sorting, where $x \in V$}} \\
    \midrule
    \textbf{Data Focus\tnote{1}} & {\makecell[c]{\textbf{Common Signatures of Fields}}} \\
    \midrule
    \makecell[c]{{Range-aware}} & {\makecell[l]{$\{f\}$, where the proportion of cells ranged between 0 \\ and 1 or between 1 and 100 is greater than 0.97.}} \\
    \midrule
    \makecell[c]{{Field Type}} & {\makecell[l]{$\{f\}$, whose field type satisfies the restrictions: user \\intent of  \textit{Time Trend} requires date field as x-axis, \\\textit{Comparison} requires string field as x-axis,  \textit{Relation} \\requires numerical field as x-axis }} \\
    \bottomrule
    \end{tabular}%
    \begin{tablenotes}  
        \footnotesize  
        \item[1] Refer to \reffig{fig:analytical_semantics} for the \textit{User Intent} and \textit{Data Focus}.
        \item[2] $\mathcal{M}$ is a 150-word vocabulary of meaningless strings according to \textit{ConFormT}. 
        \item[3] $\mathcal{E}$ is a 20-word vocabulary of empirical numbers according to \textit{ConFormT}.
    \end{tablenotes} 
  \end{threeparttable}
  \caption{Design of common signatures for cells and fields. Cells (Fields) meet these common signatures serve as the candidates of parameter (x/y-axis) selection.
  \label{tab:answer_span}}
  \vspace{-6mm}
\end{table}%

\vspace{-1mm}
\subsection{Input Features Extraction}
\label{sec:model_input}

\subsubsection{Statistical Module}
\label{sec:distribution_feature}
In table analysis and visualization tasks, the distributional information on both cell level and field level can play a decisive role when users perform actions over tabular data \cite{Wills2010AutoVisAV, Seo2005ARF}. However, the distributional features can hardly be learned and represented with pre-trained language models and tabular models. To mitigate the limitation of the current pre-trained models in terms of number representation and the unawareness of data statistics, we devise a statistical module to capture the distributional information underneath the tables.

Based on the designed data focus and expert knowledge, we extract data features at different levels of structures as shown in Table \ref{tab:field_signatures}. The inferred distributional features can hence simultaneously serve as an indicator for cell sampling and prune candidates for data referencing in follow-up steps.


\subsubsection*{\textbf{Cell Signature Design}}
\label{sec:cell_signature}
Given the cell inputs { ${C}_i = (c_1, \dots, c_n)$} of field $i$ without the header, we try to characterize the cell $k$ from raw distributional features and inferred distributional features concluded by expert knowledge as listed in Table \ref{tab:field_signatures}.


In terms of the distributional features, we measure multiple cell signatures $\{s\}$ in \reftab{tab:field_signatures} designed as follows to enable the model to be aware of distinguishing distribution semantics respectively:

\noindent $\bullet$ \textit{Frequency signatures}: including frequency count (\ie, absolute values of cell frequency), frequency ratio (\ie, the proportion of a cell value) and frequency order (\ie, the rank of a cell's frequency). 

\noindent $\bullet$ \textit{Rank signatures}: including the ascending and descending local rank within related fields and the global rank. 

\noindent $\bullet$ \textit{Range signatures}: including range positions (\ie, position of a cell in the sorted field, such as 10\% and 25\%), positions of cells in the min-max range and log range defined in \reftab{tab:field_signatures}.

In terms of expert knowledge inferred ones, we identify whether each cell belongs to the corresponding common pattern of each data focus type as described in Table \ref{tab:field_signatures}, where the common patterns are explained in Table \ref{tab:answer_span}. Based on the inferred distributional signatures, we apply a sampling strategy to sample ${C^{'}}_i \sqsubseteq {C}_i$, where cells that do not satisfy any of the 
inferred features will be filtered out. This sampling strategy can decrease the number of cells to be fed to the model and shrink the candidate pool of data referencing, thus can improve the model efficiency and performance simultaneously\footnote{All the above signatures are calculated for the numeric fields while only the applicable ones are calculated for the string fields.}. These signatures will be the cell inputs of the follow-up machine learning model as demonstrated in Figure \ref{fig:input}. 


\subsubsection*{\textbf{Field Signature Design}}
\label{sec:field_signature}

\begin{figure}
\vspace{-3mm}
	\centering
	\includegraphics[width=\columnwidth]{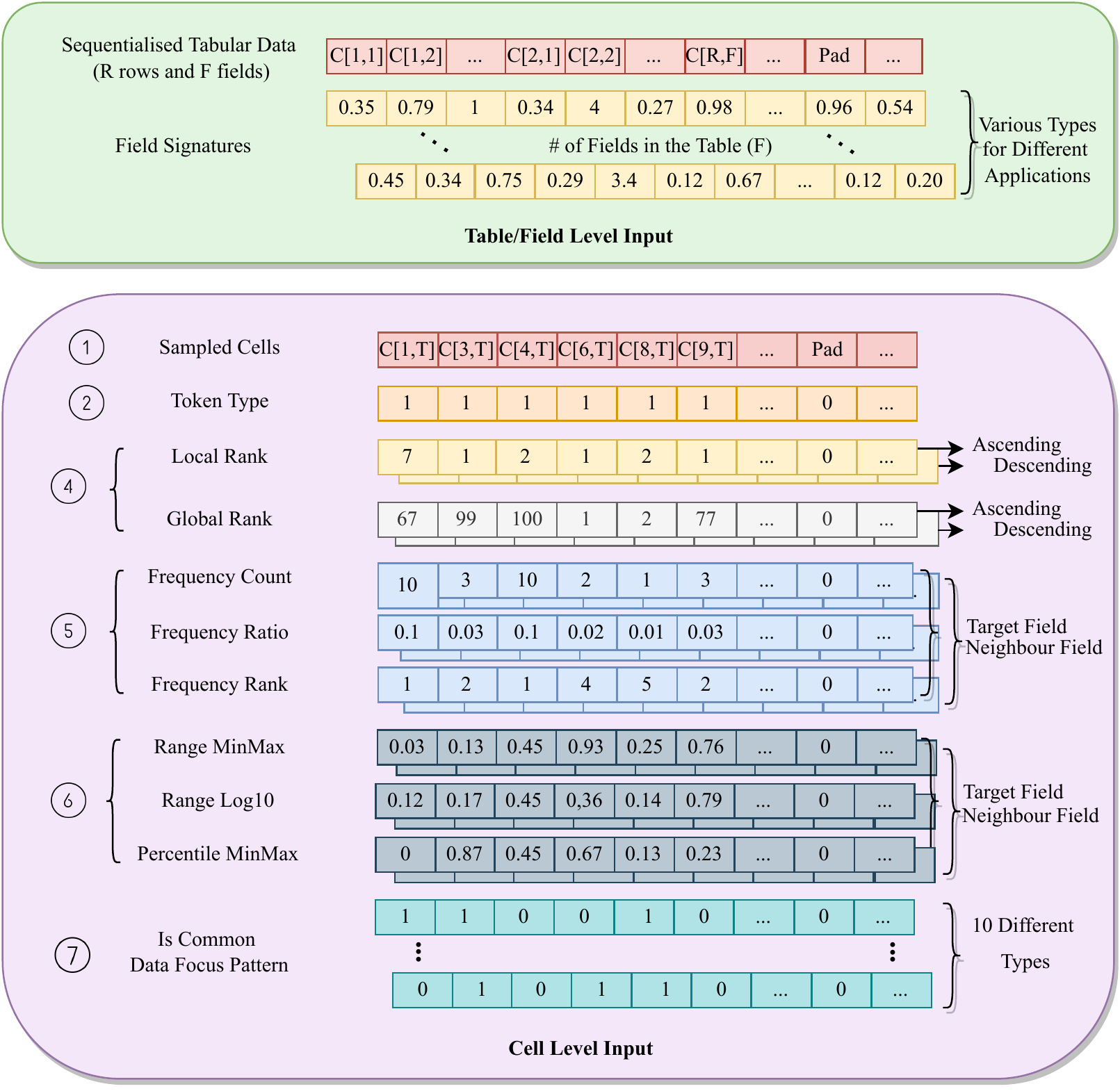}
	\vspace{-6mm}
	\caption{The input format of different structure levels for ASTA framework. \textit{R} and \textit{F} represent the number of rows and fields in the original table respectively, and \textit{T} is the index of target field when it comes to cell level input.
	\label{fig:input}} 
	\vspace{-4mm}
\end{figure}

At the field level, we extract the statistical features ${S}_{i}$ over field $i$, utilizing statistical signatures presented in Table2Analysis \cite{Zhou2020Table2AnalysisMA}, as well as more newly designed features shown in Table \ref{tab:field_signatures}. The field signatures in the lower part of \reftab{tab:field_signatures} are designed based on chart corpora; See more detailed description in \refsec{sec:chart}. These statistical signatures will be the input of the follow-up machine learning model on multi-task recommendation as the \textit{Table/Field Level Input} as shown in \reffig{fig:input}. 

\begin{table}
\footnotesize
  \centering
  \begin{threeparttable}
    \begin{tabular}{p{0.15cm}L{2.8cm}|L{4.4cm}}
    \toprule
    \multicolumn{2}{c}{\makecell[c]{\textbf{Cell Signatures}}} & \textbf{Description} \\
    \midrule
    \multicolumn{3}{l}{{Raw distributional features}} \\
    \makecell[c]{$\bullet$} & Frequency Count & Count the number of $e_i$. \\
    \makecell[c]{$\bullet$} & Frequency Ratio  & Compute the proportion of $e_i$. \\
    \makecell[c]{$\bullet$} & Frequency Rank  & Rank of $\textit{freq}(e_i)$ after sorting. \\
    \makecell[c]{$\bullet$} & Ascending Rank & Position of $e_i$ in ascending order.\\
    \makecell[c]{$\bullet$} & Descending Rank  & Position of $e_i$ in descending order. \\
    \makecell[c]{$\bullet$} & Range MinMax\tnote{1} & Position of $e_i$ in MIN-MAX range.\\
    \makecell[c]{$\bullet$} & Range LOG\tnote{2} & Position of $e_i$ in LOG range. \\
    \makecell[c]{$\bullet$} & Percentile MinMax & Percentile of $e_i$ after sorting. \\
    \midrule
    \multicolumn{3}{l}{{Inferred distributional features}} \\
    \makecell[c]{$\bullet$} & Is Common Frequency\tnote{3}  & Whether $\textit{freq}(e_i)$ is a common frequency. \\
    \makecell[c]{$\bullet$} & Is Common Rank\tnote{3}  & Whether $\textit{rank}(e_i)$ is a common rank. \\
    \makecell[c]{$\bullet$} & Is Common Range\tnote{3}  & Whether $\textit{range}(e_i)$ is a common range. \\
    \makecell[c]{$\bullet$} & Is Meaningless Value\tnote{3}  & Whether $e_i$ is a meaningless value. \\
    \makecell[c]{$\bullet$} & Is Empirical Value\tnote{3}  & Whether $e_i$ is an empirical value. \\
    \makecell[c]{$\bullet$} & Is Blank Value  & Whether $e_i$ is blank. \\
    \makecell[c]{$\bullet$} & Is Error Value  & Whether $e_i$ is an error. \\
    \toprule
    \multicolumn{2}{c}{\makecell[c]{\textbf{Field Signatures}}} & \textbf{Description} \\
    \midrule
    \multicolumn{2}{l}{{Metadata features\tnote{4}}} &
    {DataType, KeyEntropy, CharEntropy, ...} \\
    \midrule
    \multicolumn{3}{l}{{Raw distributional features}} \\
    \makecell[c]{$\bullet$} & Field Type & Field type of $f_i$. \\
    \makecell[c]{$\bullet$} & Header Similarity  & Highest similarity of $header(f_i)$. \\
    \makecell[c]{$\bullet$} & Has Keyword X \  & whether key words (\eg "categories") exist. \\
    \makecell[c]{$\bullet$} & Has Keyword Y & whether key words (\eg "values") exist. \\
    \midrule
    \multicolumn{3}{l}{{Inferred distributional features}} \\
    \makecell[c]{$\bullet$} & Is Common Cardinality  & Whether $f_i$ has high unique value ratio\tnote{5}. \\
    \makecell[c]{$\bullet$} & Is Common Range\tnote{3}  & Whether $f_i$ is a common range. \\
    \makecell[c]{$\bullet$} & Is Common Affix  & Whether $f_i$ has common prefix or suffix. \\
    \makecell[c]{$\bullet$} & Is Common Header & Whether $f_i$ has high header similarity \tnote{5}. \\
    \makecell[c]{$\bullet$} & Is Common Type\tnote{3}  & Whether $f_i$ is a common field type. \\
    \makecell[c]{$\bullet$} & Is Date Format  & Whether date/month/year info exists in $f_i$. \\
    \bottomrule
    \end{tabular}%
    \begin{tablenotes}  
        \footnotesize 
        \item[1] Calculate ($e_i$ - \textit{MIN}) / (\textit{MAX - MIN}).
        \item[2] Calculate ($e_i$ - $10^N$) / $2 \cdot 10^N$ where $N = \max(\lceil \lg |MAX| \rceil, \lceil \lg |MIN| \rceil)$.
        \item[3] Refer to Table.\ref{tab:answer_span} for the definition of \textit{common signatures and vocabulary}.
        \item[4] Apply data features from Table2Charts \cite{zhou2021table2charts}.
        \item[5] \textit{"High unique value ration"} means that the proportion of unique values exceeds the threshold (0.99). \textit{"High header similarity"} means that the header similarity between target field and others exceeds the threshold (0.61).
    \end{tablenotes} 
  \end{threeparttable}
  \caption{Design of multi-level signatures.\label{tab:field_signatures}}
  \vspace{-7mm}
\end{table}%

\begin{figure*}
\centering
\includegraphics[width=\linewidth]{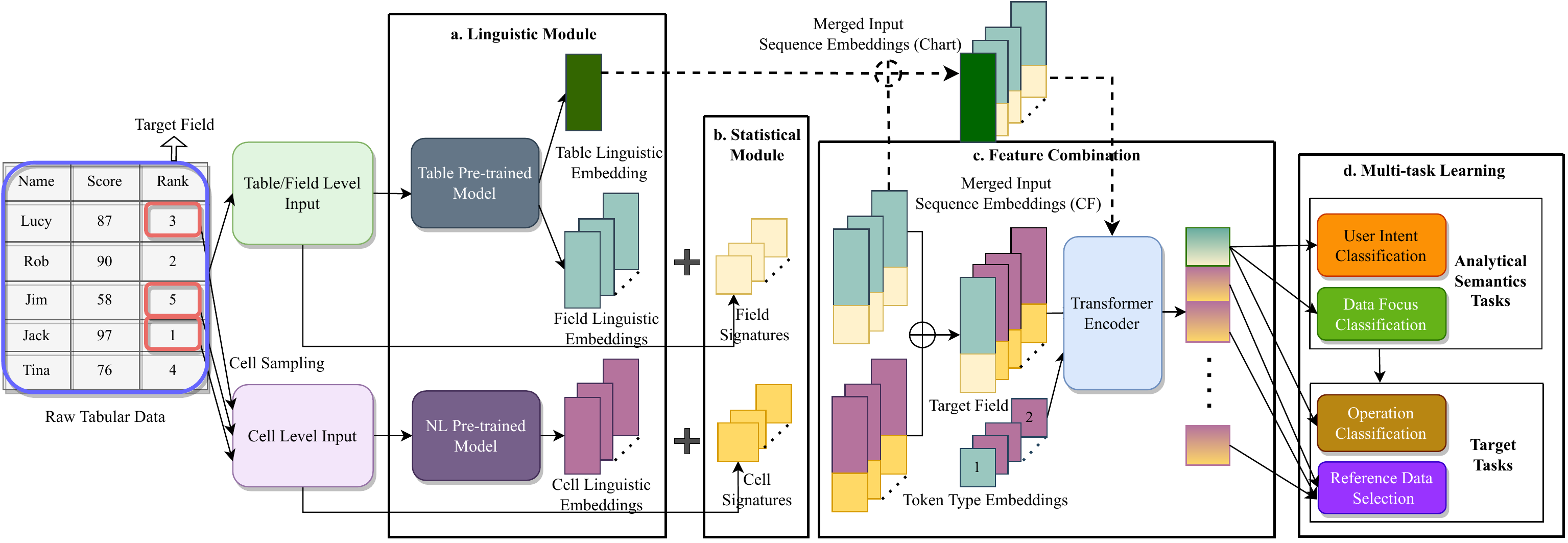}
\vspace{-5mm}
\caption{Machine learning model in ASTA framework.}
\label{fig:model}
\vspace{-3mm}
\end{figure*}

\subsubsection{Linguistic Module}
\label{sec:linguistic_feature}

It is a preliminary problem on table-related tasks that how to represent the tabular data in various down-stream tasks, as distinguishing tasks require different level of representations. The existing pre-trained tabular models \cite{herzig-etal-2020-tapas, iida-etal-2021-tabbie} can only deal with tables with constraints on table size and the maximum sentence length in a cell due to the limitation on the number of tokens entered Transformer.
In AST, we leverage diverse encoding strategies for multi-level inputs of tabular data shown in Figure \ref{fig:model}.

\subsubsection*{\textbf{Cell Linguistic Embeddings.}}
Given the sampled cells ${C^{'}}_i$ of field $i$ from \refsec{sec:cell_signature}, we feed them to the pre-trained language model (PLM) RoBERTa, where the textual or numerical contents in each cell is treated as a sentence sample and will be tokenized and encoded subsequently. As widely-used in sentence representation \cite{Reimers2019SentenceBERTSE}, the first token of the output embeddings from the PLM is conceived as the linguistic embedding of each cell. We formulate the process as $CLR_{i} = PLM({C^{'}}_i)$, where $CLR_{i} \in \mathbb{R}^{b \times e}$, $b$ is the number of cells in ${C^{'}}_i$ and $e$ is the embedding size of the pre-trained language model. Despite the fact that many pre-trained tabular models \cite{herzig-etal-2020-tapas, iida-etal-2021-tabbie} can provide cell-level representations, they can merely encode a limited number of cells based on the given queries provided in Table QA tasks, due to the constraint of token numbers in Transformer. However, the tasks without a given query, such as conditional formatting, entails cell candidates from the whole field. Therefore, we leverage pre-trained language model for each cell as their linguistic representations, and learn their contextual relationship in \refsec{sec:combine_features}. 



\subsubsection*{\textbf{Field Linguistic Embeddings.}} 
To obtain the linguistic representation $FLR_{i}$ of field $i$ that contains both general information of table context and specific features of local fields, we take advantage of pre-trained tabular model (PTM). The inputs to the PTM are all elements of a table in a sequetialized format, including cells and headers from the current field and other fields\cite{yin-etal-2020-tabert, herzig-etal-2020-tapas,iida-etal-2021-tabbie}. We extract the field embeddings from the pre-trained tabular model after feeding the serialised tabular data $T_i$ into the pre-trained model, where $i$ indicates the index of target field. To be specific, in TABBIE and TabFact, we obtain the field embedding as the [CLS] token embedding at the start of each column sequence; in TAPAS, we average the token embeddings within the same column. We formulate the process as $FLR_{i} = PTM(T_i)$, where $FLR_{i} \in \mathbb{R}^{1 \times E}$, and $E$ is the embedding size of the pre-trained tabular model.


\subsection{Machine Learning Model}
\label{sec:ml_model}

\subsubsection{Feature Combination}
\label{sec:combine_features}
After the extraction of statistical and linguistic features at different structural levels as described above, we need to transform them into the same feature space to jointly learn the target tasks.
For a specific field, we first concatenate its field signature $S_{i}$ with field embedding $FLR_{i}$, and cell signatures $\{s_{k}\}$ with their cell embeddings $CLR_{i}$ in the dimension of embedding size. Then, we transform the concatenated field representation and cell representations to the same dimension with linear layers and activation function. After that, We append the unified merged cell embedding $MCR_{i}$ after the merged field embedding $MFR_{i}$ to get the final fused embedding sequence $MSI_{i}$, which will be fed to the transformer encoder.
\begin{equation}
\begin{aligned}
&MFR_{i} = Linear_{field}(ReLU(Concate(FLR_{i}, S_{i}))) \\
&MCR_{i} = Linear_{cell}(ReLU(Concate(CLR_{i}, \{s_{k}\}))) \\
&MSI_{i} = MFR_{i} \bigoplus MCR_{i} 
\end{aligned}
\end{equation}

\noindent where $MFR_{i} \in \mathbb{R}^{D}$, $MCR_{i} \in \mathbb{R}^{b \times D}$, $MSI_{i} \in \mathbb{R}^{b \times D}$, and $D$ is the embedding size of the transformer encoder in next step.
After obtaining the fused and unified input sequences, we feed them to a Transformer encoder together with a token type embedding sequence ${TE}_i$, which indicates the different structure levels (field or cell level) of the corresponding embedding, \ie, $Q_{i} = Trans(MSI_{i}, TE_i)$. The output $Q_{i}$ is treated as the final representations of both linguistic and statistical information of the related tabular data. 

\subsubsection{Multi-Task Recommendation}
\label{sec:multi_recom}

The output embeddings from Transformer encoder will be fed to different task headers as shown in module $d$ of Figure \ref{fig:model}. We train the four tasks simultaneously, while the first two tasks as analytical semantics learning will affect the results of the target tasks from both model weights updating and their mapping relationship demonstrated in Figure \ref{fig:analytical_semantics}.


\subsubsection*{\textbf{Analytical Semantics Learning.}}
We extract the first element of $Q_i$ as the input of user intent classifier and data focus classifier, when it comes to the application of conditional formatting that is exerted on the field level. The field representation embedding $Q_i[0]$ is transformed into logits $\hat{y}_{i,u}$ and $\hat{y}_{i,d}$ after linear layers and activation function, indicating the probability of being predicted as different types of user intent and data focus. The loss for the two tasks are calculated as $\mathcal{J}_u = -w_{i,u}[p_{u}y_{i,u}\cdot\log(\sigma(\hat{y}_{i,u})) + (1 - y_{i,u})\cdot\log(1 - \sigma(\hat{y}_{i,u}))]$ and $\mathcal{J}_d = -w_{i,d}[p_{d}y_{i,d}\cdot\log(\sigma(\hat{y}_{i,d})) + (1 - y_{i,d})\cdot\log(1 - \sigma(\hat{y}_{i,d}))]$, which is utilized to optimize the model weights.


\subsubsection*{\textbf{Operation Classification.}}
We further recommend the possible operation types of each field as a multi-label classification task. The field embedding $Q_i[0]$ is fed to separate linear layers and outputs $\hat{y}_{i,o}$ as probabilities of recommendation for each operation type. The loss ${J}_{o}$ is formulated as $\mathcal{J}_{o} = -w_{i,o}[(p_{o}y_{i,o}\cdot\log(\sigma(\hat{y}_{i,o}))
y_{i,o})\cdot\log(1 - \sigma(\hat{y}_{i,o})))]$.
It is worth noting that, the evaluation of operation task will be affected by the predicted results of analytical semantics, which is served as two auxiliary intuitive tasks that can guide the more complicated analysis tasks. Specifically, the predicted types of user intent and data focus will filter out the candidates of operation type as shown in Figure \ref{fig:analytical_semantics}, which specifies what operations should not appear based on the combination of predicted analytical semantics, thus shrinking the decision space.

\subsubsection*{\textbf{Reference Generation.}}
After the recommendation of operation type, our model recommends the most probable parameter candidates for the recommended operation type. The data referencing is achieved by utilizing the last hidden layer embeddings $Q_{i}[1:]$ from the Transformer encoder as the cell representation embeddings, and passing them to the referenced data selector. In the selector, the representation embedding of each cell will be fed into linear layers, which output the probabilities of cells being recommended for the corresponding operation types. The process will be formulated as a multi-label classification problem with an optimization loss $\mathcal{J}_{p}$ defined similarly to $\mathcal{J}_{o}$. In particular, credited to the inferred distributional features we designed in Table \ref{tab:field_signatures}, ASTA can simply focus on the cells that satisfy the common pattern corresponding to the predicted data focus, and thus pruning the parameter candidates.

The final aggregated loss is then defined as $\mathcal{J}_{final} = \alpha\mathcal{J}_u + \beta\mathcal{J}_{d} + \gamma\mathcal{J}_{o} + \delta\mathcal{J}_{p}$, where $\alpha$, $\beta$, $\gamma$ and $\delta$ are the scaling coefficients.

\subsection{Application on Chart Recommendation}
\label{sec:chart}


We utilize chart recommendation to demonstrate how our ASTA framework can be applied to a broader range of table analysis and visualization in a similar way as described above.
Charts are common choices for users to visualize tabular data~\cite{hu2019viznet,dibia2019data2vis,hu2019vizml}.
Compared with the cell-level parameter generation in conditional formatting, chart creation requires selecting table fields as x/y-axis~\cite{zhou2021table2charts} and is therefore the field-level referencing in a table.

\subsubsection{Chart Recommendation.} 
Referring to the existing works, we standardize chart creation to answer two essential questions: 1) Which chart type is chosen to visualize data? 2) Which fields are referenced as axes? In this sense, we define the action space of chart creation into the selection of chart type $\mathcal{O}_{\textit{Chart}}$ and x/y-axis $\mathcal{R}_{\textit{Chart}}$.

\begin{myDef}
(Chart) An executable chart record on table $T$ requires both a chart type $o$ and several optional axes $\{r\}$:
\begin{equation}
\begin{aligned}
Chart(T):= \{ o, r_i, \cdots 
&\mid o \in \mathcal{O}_{Chart}, r \in \mathcal{R}_{Chart}, i \in \mathbb{N}^+\}
\end{aligned}
\end{equation}
\end{myDef}

\noindent $\bullet$ \textit{Chart Type ($\mathcal{O}_{Chart}$).} We focus on four major chart types in public chart corpora published in \cite{hu2019vizml, zhou2021table2charts}, including \textit{"Bar chart"}, \textit{"Line chart"}, \textit{"Scatter chart"} and \textit{"Pie chart"}. To show the effects of different charts, we provide two examples in \reffig{fig:cf_usage}(c) and \ref{fig:cf_usage}(d).

\noindent $\bullet$ \textit{X/Y-Axis ($\mathcal{R}_{Chart}$).} A chart requires at least one y-axis $r^y$, while the x-axis $r^x$ is optional according to the chart type.\footnote{For example, creating a bar chart requires an x-axis, while a pie chart does not.}.

\subsubsection{User Intent and Data Focus.}
\label{sec:chart_analytical_semantics }


Given the public chart corpora~\cite{hu2019vizml,zhou2021table2charts} and related studies~\cite{szoka1982guide,wilke2019fundamentals}, we design the analytical semantics of chart with the same steps from \refsec{sec:data_focus} as follows.

\noindent $\bullet$ \textit{User Intent ($\mathcal{U}_{Chart}$).}
we summarize four user intents for charts.
\begin{equation}
\begin{aligned}
\mathcal{U}_{Chart} := \{ \textit{Rlt}, \textit{Cps}, \textit{Cpr},\textit{Ttr} \} 
\end{aligned}
\label{eq:user_chart}
\end{equation}
\textit{Rlt} emphasizes the relationship between variables, \textit{Cps} describes a part of the whole, \textit{Cpr} compares multiple things, and \textit{Ttr} shows a time series. There is a many-to-many mapping between user intent and chart type, and user intents can exist simultaneously on a chart.

\noindent $\bullet$ \textit{Data Focus ($\mathcal{D}_{Chart}$).} Based on the characteristics of charts, we present six categories to highlight different data features, \ie, 
\begin{equation}
\begin{aligned}
\mathcal{D}_{Chart} := \{ \textit{Fmt}, \textit{Caf}, \textit{Hsi}, \textit{Rag}, \textit{Car}, \textit{Fty} \} 
\end{aligned}
\label{eq:data_chart}
\end{equation}
They represent field type (\textit{Fty}), cardinality (\textit{Car}), range (\textit{Rag}), common affix (\textit{Caf}), header similarity (\textit{Hsi}) and date format (\textit{Fmt}), with some common signatures similarly designed in \reftab{tab:answer_span}. These data focuses depict diverse aspects data characteristics of each table field.
Note that there are differences in crucial features for the horizontal and vertical axes: $\mathcal{D}^{X}_{\textit{Chart}} = \{\textit{Fmt}, \textit{Caf}, \textit{Hsi}, \textit{Fre}, \textit{Fty}\}$ for x-axis and $\mathcal{D}^{Y}_{\textit{Chart}} = \{ \textit{Hsi}, \textit{Rag}, \textit{Fre}, \textit{Fty}\}$ for y-axis.

\noindent $\bullet$ \textit{Machine Learning Tasks.} The machine learning tasks for chart recommendation include analytical semantics learning and several chart tasks, \ie, chart type classification, x-axis and y-axis generation (including x/y-axis number decision tasks, x/y-axis fields selection tasks) and complete chart recommendation task.


\subsubsection{Input Features and Machine Learning Model.}

Similar to conditional formatting involving field and cell representations, charts also require features from distinct levels of table structure. The difference is that chart recommendation chooses chart type and x/y-axis based on the representation of overall table and field respectively, while the cell representation is not necessary.
In this case, we only enter the ASTA framework from the \textit{Table/Field Level Input} flow in \reffig{fig:input} and \ref{fig:model}. The linguistic module produces table-level and field-level linguistic embeddings. In statistical module, we extract the chart-related field signatures in Table \ref{tab:field_signatures} to combine it with field linguistic embeddings. Note that we have not designed table-level statistical signatures and leave it as future work.

The following steps are similar to those in conditional formatting tasks. The table-level merged representation generated by Transformer encoder will be fed to the the first three headers in \reffig{fig:model} simultaneously, while the field-level representations will be input to the last header -- \textit{referenced data Selection} header. 
Similar applications of ASTA framework on other table analysis and visualization can be performed by the above method, involving customised pre-trained language and tabular model, and self-defined statistical signatures at different levels.

\section{Experiments}
\label{sec:experiments}

Our ASTA framework is evaluated on analytical semantics learning and follow-up tasks, including conditional formatting (\refsec{sec:cf_result}) and chart recommendation (\refsec{sec:chart_result}), via newly collected corpus ConFormT and public chart corpora. We also verify the reasonability of ASTA framework by ablation study and human evaluation in \refsec{sec:ablation_human}.

\subsection{General Setup}
\label{sec:setup}

We choose the pretrained XLM-RoBERTa-base model \cite{conneau-etal-2020-unsupervised} with L = 12, H = 768, A = 12 and 270M parameters as the pretrained language model in ASTA framework.
When feeding the word sequence within a cell or header into the pretained language model, we only keep the first 30 tokens. In feature combination step, the Transformer encoder is set as $H_{enc} = 256$, $A = 8$ and $L = 6$. 
The experiments are run on Linux machines with 24 CPUs, 448GB memory and 4 NVIDIA Tesla V100 16G-memory GPUs. Although we use 4 GPUs for each training, all evaluations are done on 1 GPU for fair comparisons with the same configuration. And we set the loss weights $\alpha$, $\beta$, $\gamma$ and $\delta$ all equal to 1 in both tasks.
The corpus is randomly allocated for training, validation and testing in the ratio of 7:1:2. We show the results on test set in the following sections.



\begin{table}
  \centering
  \footnotesize
  \begin{threeparttable}
    \begin{tabular}{c|c|ccc}
    \toprule
    \multicolumn{2}{c|}{\multirow{2}{*}{\makecell[c]{\textbf{Conditional}  \textbf{Formatting}}}} & \multirow{2}{*}{\makecell[c]{${\textbf{AST}_{TABBIE}}$}} & \multirow{2}{*}{\makecell[c]{${\textbf{AST}_{TAPAS}}$}}  & \multirow{2}{*}{\makecell[c]{${\textbf{AST}_{TabFact}}$}}   \\
    \multicolumn{2}{c|}{} & & & \\
    \midrule
    \multirow{2}{*}{\makecell[c]{Overall \\ ($CF$)}} & R@1  & \textbf{72.31} & 67.40 & 68.36  \\
    & R@3  & \textbf{85.25} & 81.60 & 82.44  \\
    \midrule
    \multirow{2}{*}{\makecell[c]{Operation \\ ($\mathcal{O}_{\textit{CF}}$)}} & R@1  & \textbf{81.00} & 77.06 & 78.70  \\
    & R@3  & \textbf{92.10} & 90.16 & {90.88} \\
    \midrule
    \multirow{2}{*}{\makecell[c]{Parameter \\ ($\mathcal{R}_{\textit{CF}}$)}} & R@1  & \textbf{82.74} & 79.46 & 79.30 \\
    & R@3  & \textbf{90.79} & 88.29 & 88.91  \\
    \midrule
    Equal & \multirow{12}{*}{R@1} & \textbf{78.26} & 69.19 & 73.78 \\
    L/G Than & & \textbf{72.86} & 72.31 & 70.62 \\
    Top K & & \textbf{69.82} & 47.69 & 38.55 \\
    Between & & \textbf{47.31} & 43.18 & 41.72 \\
    Is Duplicate & & \textbf{27.94} & 1.11 & 5.00 \\
    Is Blank & & {77.03} & \textbf{81.73} & 79.23 \\
    Is Error & & {95.78} & \textbf{96.96} & 95.71 \\
    Equal Set & & {84.56} & 84.78 & \textbf{89.18} \\
    Partition Set & & \textbf{65.88} & 57.41 & 59.90 \\
    Color Scale & & \textbf{89.72} & 85.53 & 86.46 \\
    Data Bar & & {67.50} & 59.94 & \textbf{68.64} \\
    Icon Set & & {67.21} & \textbf{74.38} & 61.22 \\
    \bottomrule
    \end{tabular}%
  \end{threeparttable}
  \caption{Evaluation on conditional formatting recommendation (averaged over 3 runs): (Top) Recall numbers of complete task (\ie, \textit{overall}) and sub-tasks (\ie, \textit{operation} and \textit{parameter}). (Bottom) Overall recall of each operation type.
  \label{tab:eval_cf}}
  \vspace{-7mm}
\end{table}%


\begin{table*}
  \centering
  \footnotesize
  \begin{threeparttable}
    \begin{tabular}{c|p{2cm}|r|c|c|rr|rr|rr|rr}
    \toprule
    \multicolumn{2}{c|}{\multirow{2}{*}{\textbf{Chart}}}   & \multirow{2}{*}{\makecell[c]{\textbf{Overall} \\ ($Chart$)}} &  \multirow{2}{*}{\makecell[c]{\textbf{Chart Type} \\ ($\mathcal{O}_{\textit{Chart}}$)}} & \multirow{2}{*}{\makecell[c]{\textbf{X/Y-Axis} \\ ($\mathcal{R}_{\textit{Chart}}$)}} & \multicolumn{2}{c|}{\multirow{2}{*}{\textbf{Line}}} & \multicolumn{2}{c|}{\multirow{2}{*}{\textbf{Bar}}} &
    \multicolumn{2}{c|}{\multirow{2}{*}{\textbf{Scatter}}} & \multicolumn{2}{c}{\multirow{2}{*}{\textbf{Pie}}} \\
    \multicolumn{2}{c|}{} &&&&&&&&&&& \\
    \midrule
    {Corpus} & \makecell[c]{{Model}}  & \makecell[c]{{R@1}} & \makecell[c]{{R@1}} & \makecell[c]{{R@1}} & {R@1} & {P@1} & {R@1} & {P@1}& {R@1} & {P@1} & {R@1} & {P@1} \\
    \midrule
    \multirow{4}{*}{\makecell[c]{{Excel}}} & \makecell[c]{${{AST}_{TABBIE}}$} & \textbf{54.41} & \textbf{73.29} & \textbf{78.14} & \textbf{43.11} & \textbf{46.71} & \textbf{45.16} & 52.04 & \textbf{66.77} & \textbf{65.68} & \textbf{73.86} & \textbf{51.18} \\
    & \makecell[c]{{DeepEye}} & 15.73 & 40.11 & 32.13 & 1.63 & 13.62 & 16.24 & 16.28 & 24.46 & 9.07 & 0.54 & 22.39 \\
    & \makecell[c]{{Data2Vis}} & 7.36 & 47.79 & 14.43 & 3.38 & 3.62 & 6.97 & 12.22 & 6.05 & 4.45 & 3.91 & 5.83 \\
    & \makecell[c]{{Table2Charts}} & 42.88 & 68.74 & 55.86 & 42.10 & 36.02 & 43.57 & 60.03 & 28.87 & 35.14 & 60.58 & 51.17 \\
    \midrule
    \multirow{4}{*}{\makecell[c]{{Plotly}}} & \makecell[c]{${{AST}_{TABBIE}}$} & \textbf{62.86} & \textbf{82.60} & {74.57} & \textbf{93.58} & \textbf{95.42} & \textbf{52.64} & {50.84} & {49.52} & \textbf{47.70} & \textbf{98.19} & \textbf{96.29} \\
    & \makecell[c]{{DeepEye}} & 25.06 & 36.51 & 36.51 & 24.15 & 34.00 & 3.42 & 6.43 & 22.78 & 12.01 & 0.00 & 0.00 \\
    & \makecell[c]{{Data2Vis}} & 23.11 & 33.14 & 69.25 & 10.95 & 10.21 & 10.15 & 34.44 & 26.96 & 12.60 & 0.56 & 1.30 \\
    & \makecell[c]{{Table2Charts}} &  48.64 & 49.48 & 88.03 & 91.21 & 68.39 & 5.80 & 87.52 & 90.61 & 24.97 & 98.00 & 67.02  \\
    \bottomrule
    \end{tabular}%
  \end{threeparttable}
  \caption{Evaluation of chart recommendation (averaged over 3 runs): (Left) Recall numbers of complete task (\ie, \textit{overall}) and sub-tasks (\textit{chart type} and \textit{x/y-axis}). (Right) Overall recall and precision of each chart type.
  \label{tab:eval_chart}}
  \vspace{-7mm}
\end{table*}

\begin{table}
  \centering
  \footnotesize
  \begin{threeparttable}
    \begin{tabular}{l|c|ccc}
    \toprule
    \multirow{2}{*}{\makecell[c]{\textbf{Model}}}  & \multirow{2}{*}{\makecell[c]{\textbf{Corpus}}} &  \multirow{2}{*}{\makecell[c]{\textbf{Overall} \\ ($AS$)}} &
    \multirow{2}{*}{\makecell[c]{\textbf{User Intent} \\ ($\mathcal{U}$)}} & \multirow{2}{*}{\makecell[c]{\textbf{Data Focus} \\ ($\mathcal{D}$)}} \\
    & & & & \\
    \midrule
    ${\textbf{AST}_{TABBIE}}$ & \multirow{3}{*}{ConFormT} & 85.88 & 88.63 & 96.13 \\
    ${\textbf{AST}_{TAPAS}}$ & & 80.05 & 82.94 & 94.52 \\
    ${\textbf{AST}_{TabFact}}$ & & 82.35 & 85.84 & 93.77 \\
    \midrule
    \multirow{2}{*}{${\textbf{AST}_{TABBIE}}$} & Excel Chart & 88.19 & 90.89 & 98.07 \\
    & Plotly Chart & 94.53 & 97.09 & 98.55 \\
    \bottomrule
    \end{tabular}%
  \end{threeparttable}
  \caption{Evaluation on analytical semantics learning of conditional formatting and chart (averaged over 3 runs): Recall at top 1 of complete analytical semantics and sub-tasks (\ie, \textit{user intent} and \textit{data focus}).
  \label{tab:eval_as}}
  \vspace{-9mm}
\end{table}%

\subsection{Conditional Formatting Recommendation}
\label{sec:cf_result}

\subsubsection{Baselines.}
\label{sec:baseline}
There is no previous model that can perfectly handle conditional formatting recommendation. Some semantic parsing models on tabular data are able to provide cell embeddings that can be used for cell selection, \eg TAPAS~\cite{herzig-etal-2020-tapas}, TaBERT~\cite{yin-etal-2020-tabert}, RNNCell\cite{icdm2019}, TABBIE and TabFact~\cite{2019TabFact,chen2020open}. 
Given that RNNCell treats numbers as unknown tokens,
this model cannot handle references to numbers in this problem. 
\textbf{TAPAS}~\cite{herzig-etal-2020-tapas} and \textbf{TaBERT}~\cite{yin-etal-2020-tabert} both provide pretrained learning models that jointly learn representations for natural language sentences and tables (targeting table question answering task). 
\textbf{TabFact}~\cite{2019TabFact} proposes a TABLE-BERT model and encodes sub-table with natural language templates.
\textbf{TABBIE}~\cite{xiang:2020:TURL} learns deep contextualized representations on relational tables using pre-training/fine-tuning paradigm.
It proposes a structure-aware Transformer encoder to model the row-column structure of relational tables. 
In summary, we utilize TAPAS, TabFact and TABBIE models in the ASTA framework to generate field representations and compare their performance.

\subsubsection{Evaluation Metrics.}
For the step-by-step tasks and complete recommendation task, recall at top-k (R@k, k=1,3) numbers are adopted as evaluation metrics. 

In the evaluation of recall at top-k in \textit{Overall (CF)}, \textit{Operation ($\mathcal{O}_{CF}$)} and \textit{Parameter ($\mathcal{R}_{CF}$)} tasks, a table field is successfully recalled if there is at least one true answer created by users in the top k predicted records, operations or parameters, respectively. As shown in Table \reftab{tab:eval_cf}, we calculate the percentage of table fields successfully recalled among all fields, \ie, 
$$
R @ k=\frac{\#(\text{Table fields successfully recalled})}{\#(\text{Table fields)}}
$$

\noindent Further, the evaluation for each operation type is calculated as 
$$
R @ 1 =\frac{\#(\text{Table fields with target operation successfully recalled})}{\#(\text{Table fields with target operation)}}
$$
For each target operation (\eg, \textit{"Equal"}), we calculate the proportion of table fields that are successfully recalled in all the fields to which conditional formatting of target operation is applied.

\subsubsection{Results.}
We generate a top-k recommendation list of operations and parameters to complete conditional formatting records. 
Recall at top-k numbers in \reftab{tab:eval_cf} reveal how well the predicted top-k conditional formatting records by each model match the user-created records. We compare three models that use ASTA framework but are based on different pre-train models, \ie, TABBIE, TAPAS, and the model proposed in TabFact work. The model performance is demonstrated at each stage (top) and for each operation (bottom).
It shows that ASTA framework based on TABBIE model performs best on most tasks, which achieves 72.31\% (or 85.25\%) in overall Recall at top 1 (or 3) and outperforms the other two models by about 5 points on average. The advantages of $AST_{TABBIE}$ can also be seen in the tasks of operation selection (R@1 81.00\%) and parameter generation (R@1 82.74\%). We attribute this to the fact that TABBIE model is pre-trained exclusively using tabular data and provides embeddings of all table substructures, allowing the model to produce a stronger sense of table fields and the associations between fields. This effectively facilitates the recommendation of conditional formatting that is applied to table fields.

In terms of the recall for each operation type, ASTA framework based on TABBIE model also holds the lead on seven of the twelve tasks. In addition, the good performance of $AST_{TabFact}$ on \textit{"Equal Set"} and \textit{"Data Bar"} shows its ability in multi-parameter recommendation, indicating the strength of TabFact model in mining the correlation between cells. 
In contrast, the high recall values of $AST_{TAPAS}$ on \textit{"Is Blank"} and \textit{"Is Error"} show that it performs well in the fields containing special values such as error and blank. This demonstrates the potential of TAPAS model in differentiating cells.

\subsection{Chart Recommendation}
\label{sec:chart_result}

\subsubsection{Baselines.}
In the chart recommendation, data-driven approaches are becoming popular in recent learning-based systems such as DeepEye~\cite{luo2018deepeye}, Data2Vis~\cite{dibia2019data2vis} and Table2Charts~\cite{zhou2021table2charts}. 
\textbf{DeepEye}~\cite{luo2018deepeye} provides two public models (ML and rule-based) without training scripts. The ML uses a supervised learning-to-rank model to rank charts and works better.
\textbf{Data2Vis}~\cite{dibia2019data2vis} formulates visualization generation as a language translation problem, where data specifications are mapped to visualization specifications in a declarative language (Vega-Lite).
\textbf{Table2Charts}~\cite{zhou2021table2charts} does table-to-sequence generation based on deep Q-learning with copying mechanism and heuristic searching, and learn a shared representation of table fields.

\subsubsection{Evaluation Metrics.}
Similar to conditional formatting, we apply recall at top-k (R@k, k=1,3) values as evaluation metrics. A table is successfully recalled if at least one predicted chart type, referenced field set and their combination match the true answers. We calculate the recall at top k (R@k) in \textit{Chart type} ($\mathcal{O_\textit{Chart}}$), \textit{X/Y-axis} ($\mathcal{R_\textit{Chart}}$) or \textit{Overall} (\textit{Chart}) tasks respectively, as follows
$$
R @ k=\frac{\#(\text{Tables successfully recalled})}{\#(\text{Tables)}}
$$

The evaluation for each chart type includes the top 1 recall and precision values. The recall at top 1 is calculated as 
$$
R @ 1 =\frac{\#(\text{Tables with target chart type successfully recalled})}{\#(\text{Tables with target chart type)}}
$$
The precision at top 1 is calculated as 
$$
P @ 1 =\frac{\#(\text{Tables with predicted chart type successfully recalled})}{\#(\text{Tables with predicted chart type)}}
$$
In summary, the denominators of R@1 and P@1 are tables with target charts created by users, or tables for which target charts are predicted to be constructed, respectively. 
They differ from the evaluation metrics used in the baseline works, and we recalculate the values defined above for each model.

\subsubsection{Results.}
~\reftab{tab:eval_chart} shows the results of ASTA framework and baseline models on chart recommendation tasks. The evaluations on two public chart corpora (\ie, Excel and Plotly) present the outstanding performance of ASTA framework (exceeding baselines at least 12 points in the overall recall number), especially on Excel corpus. Given that users may create several charts for one table, we provide both recall and precision numbers as evaluations on each chart type. We can see that ASTA framework more or less surpasses other models on most tasks with a balanced recall and precision numbers, \eg, $R@1 = 43.11$ and $P@1 = 46.71$ in \textit{"Line"} task. ASTA also shows a stronger advantage on minor chart types, \ie, \textit{"Scatter"} and \textit{"Pie"}.
These indicate that ASTA framework can handle the recommendation of minor chart types. It should be noted that the recall and precision of some baseline models show a big difference, such as Table2Charts model on \textit{"Bar chart"} tasks. We investigate the reasons, namely that Table2Charts model focuses on major chart types and generate very few bar charts, leading to a high precision and low recall value.

\subsection{Ablation Study and Human Evaluation}
\label{sec:ablation_human}


\subsubsection{Ablation Study.}
We evaluate ASTA framework on learning analytical semantics of conditional formatting and chart. Results in \reftab{tab:eval_as} show that both on the recommendation of conditional formatting and chart, ASTA framework learned the analytical semantics well, with all recall at top 1 numbers greater than 80\%. In addition, ASTA framework based on TABBIE model outperforms other pre-trained models by at least 3\% on the \textit{Overall (AS)}, \textit{User Intent ($\mathcal{U}$)} and \textit{Data Focus ($\mathcal{D}$)} tasks.

\reftab{tab:eval_model} shows the results of model ablations, where the evaluation metrics are recall at top-1 numbers. We test the utility of analytical semantics paradigm, as well as the statistical and linguistic modules in ASTA framework. 

\noindent $\bullet$ \textit{Without analytical semantics.} We remove the module that learns and generates user intent and data focus from ASTA framework and instead make recommendations directly on conditional formatting operations and parameters. 
\reftab{tab:eval_model} show that the our design that explicitly extracts user intent and data focus effectively improves the recommendation results, \eg, by 24\% on overall conditional formatting (\textit{CF}) task and by 21\% on data referencing ($\mathcal{R}_{CF}$) task.

\noindent $\bullet$ \textit{Without statistical module.} To eliminate the distribution information, we remove the statistical module from the input of our framework. 
The results show that statistical information can improve the performance by at least 2 percentage points.

\noindent $\bullet$ \textit{Without linguistic module.} We throw away the field header and cell embedding vectors from the model input to test the function of linguistic information. The results likewise demonstrate the effectiveness of this design. In particular, without linguistic module, the recall on analytical semantics learning (\textit{AS}) is reduced by 12\%.


\subsubsection{Human Evaluation.}
The above results are based on golden labels created by rules in \refsec{sec:analytical_semantics}. 
To verify the rationality of our design and rules for analytical semantics, we compare manual labels given by experts and our rule-based labels in human evaluation.

To perform human evaluation, we randomly collect 600 unique HTML tables crawled from the public web. We create questionnaires to invite users to manually annotate analytical semantics of the given table. Each questionnaire contains one table and several multiple-choice questions about user intent and data focus. Five experts working on table visualization are required to answer each questionnaire as he or she sees fit and thus generates manual labels for the given table. We compare the manual labels with the rule-based golden labels used in experiments and calculate the recall values shown in \reftab{tab:eval_human}. The recall values are calculated as follows
$$Recall = \frac{\#(\text{Tables with golden labels successfully recalled})}{\#(\text{Tables)}}$$
Here, a table is successfully recalled if at least one rule-based label appears in the corresponding manual labels.

Recall values in \reftab{tab:eval_human} show that there is a high similarity (recall > 75\%) between rule-based and labels created by users, which verify the rationality of our design and rules for user intent and data focus. This further justifies the our findings in the above experiments.

\begin{table}
  \centering
  \footnotesize
  \begin{threeparttable}
    \begin{tabular}{p{3.8cm}|cccc}
    \toprule
    \makecell[c]{\textbf{Ablation Study
    }} & $AS$ \tnote{1} & $CF$ & $\mathcal{O}_{\textit{CF}}$ & $\mathcal{R}_{\textit{CF}}$ \\
    \midrule
    \textbf{AST} &  \textbf{85.88} & \textbf{72.31} & \textbf{81.00} &  \textbf{82.74} \\
    \quad w/o analytical semantics & \makecell[c]{-} & \makecell[c]{48.24} & \makecell[c]{57.36} & \makecell[c]{61.16}  \\
    \quad w/o statistical input module & \makecell[c]{83.92} & \makecell[c]{64.35} & \makecell[c]{77.56} & \makecell[c]{75.08}  \\
    \quad w/o linguistic input module & \makecell[c]{77.70} & \makecell[c]{65.12} & \makecell[c]{75.58} & \makecell[c]{78.63}  \\
    \bottomrule
    \end{tabular}%
    \begin{tablenotes}  
        \small  
        \item[1] "AS" (overall R@1 in analytical semantics task), "CF"  (overall R@1 in conditional formatting task), "$\mathcal{O}_{\textit{CF}}$" and "$\mathcal{R}_{\textit{CF}}$" (R@1 in operation task and parameter task of conditional formatting).
    \end{tablenotes} 
    \end{threeparttable}
  \caption{Performance of different model modifications on analytical semantics learning and conditional formatting recommendation.} \label{tab:eval_model}
  \vspace{-6mm}
\end{table}%

\begin{table}
  \centering
  \footnotesize
  \begin{threeparttable}
    \begin{tabular}{c|c|p{1.68cm}p{1.66cm}}
    \toprule
    \makecell[c]{\textbf{Human Evaluation}} & \textbf{User Intent} & \multicolumn{2}{c}{\textbf{Data Focus}} \\
    \midrule
    Conditional Formatting & 76.92\%  & \multicolumn{2}{c}{93.31\%}  \\
    \midrule
    Chart & 79.67\% & 85.67\% (X-axis)  & 96.33\% (Y-axis) \\
    \bottomrule
    \end{tabular}%
    \end{threeparttable}
  \caption{Human evaluation on analytical semantics of conditional formatting and chart: comparison between manual annotation and rule-based labels.} \label{tab:eval_human}
  \vspace{-8mm}
\end{table}%


\subsubsection{Case study.} 
To show what analytical semantics that ASTA framework learns, we randomly observe 150 tables from the test set and summarise some patterns, with examples shown in \reffig{fig:cf_usage}.

\noindent $\bullet$ Field $f^{1a}_2$: The pattern is \textit{(numerical field of competition results $\to$ highlight top 3 records)}. It shows that users tend to filter out the top 3 records to generate award candidates.

\noindent $\bullet$ Field $f^{1a}_3$: The pattern is \textit{(string field with meaningless text $\to$ highlight the meaningful records)}. It indicates that users pay attention to the part of the data that convey useful information.


\noindent $\bullet$ Field $f^{1b}_3$: The pattern is \textit{(numerical field of investment $\to$ compare the positive and negative returns)}. It suggests that users are concerned about about generating revenue or not.

In addition to the examples in \reffig{fig:cf_usage}, there are many patterns that cannot be fully enumerated.
This is why we provide the model with signatures related to analytical semantics and design tasks on user intent and data focus to learn the motivation behind table analysis and visualization implicitly through machine learning models, instead of directly labeling analytical semantics of tabular data through heuristic modules. 

\section{Related Work}
\label{sec:related}

\noindent {\bfseries Analysis and Visualization Recommendation}
Automatic recommendation of visual and analytical tasks is becoming more and more prominent in data analysis tools~\cite{milo2020overview, polaris2002}, \eg, data preparation~\cite{yan2020autosuggest}, pivot table~\cite{Zhou:2020:Table2Analysis} and chart~\cite{zhou:2021:table2charts} recommendations are widely studied topics in this area. 
However, conditional formatting~\cite{Abramovich04spreadsheetconditional} of table fields did not draw much attention from the research community, despite its importance in major tabular analysis tools from Excel, Google Sheets to Python Pandas.

In order to make recommendations according to the large-scale historical data, the previous works primarily fall in two types of methods, which are rule-based and machine learning-based. The rule-based methods leverage heuristic rules obtained from experiments or existing theories to rank the choices of visualization. Voyager \cite{Wongsuphasawat2016VoyagerEA} ranks visualizations based on both data properties and perceptual principles. 
SEEDB \cite{Vartak2015SeeDBED} prunes view spaces for aggregations by data distributions including variance, correlation and frequency. VizDeck and Foresight \cite{Key2012VizDeckSD, Demiralp2017ForesightRV} also facilitate visual insights discovery with distribution-aware algorithms. However, a limitation of the existing rule-based methods is that the expert knowledge learned is static and requires significant manual effort to derive, making them too brittle for applications.

While the machine learning-based works devise models driven by large amount of dataset to learn the design choices of visualization. 
DeepEye \cite{luo2018deepeye} leverages a supervised learning-to-rank model to rank visualizations, combining expert rules. Data2Vis \cite{dibia2019data2vis} exploits multilayered attention-based LSTM to map data to visualization specifications in Vega-Lite language.  Table2chart \cite{zhou:2021:table2charts} learns common patterns in collected corpus based on Q-learning and heuristic searching to generate chart templates. Nevertheless, these methods rarely reveal the causes for the recommended results.

In addition, despite the reasonable solutions they provide, few of them can handle cell referencing in conditional formatting task. Our framework, making up for the disadvantages above, can provide explainable recommendations and generalise to visual recommendation tasks at cell, field and table level with the backbone of pre-trained table and language models. 



\noindent {\bfseries User Intent Abstraction}
To make more explainable and accurate recommendations, there are a series of works focusing on abstracting user intent from visual and analytical operations \cite{Tessera2021,mulleveltypo2013,provenance2008}.  \cite{mulleveltypo2013} describes visualization tasks with a multi-level typology, shrinking the gap between high-level analysis and low-level tasks. \cite{provenance2008} derives a multi-tier characterization of user visual analytic activity based on Activity Theory \cite{actTheo} to extract insight provenance. Tessera \cite{Tessera2021} segments analyst event logs with interactions, data and user features to create discrete blocks of goal-directed activity and capture user changing goals dynamically. Compared to the existing works, we further differentiate data focus from the general user intents, revealing the potential motivations of the user intents driven by the distribution or the semantics of the data.

\noindent{\bfseries Table Representation}
The foundation of table understanding relies heavily on the representation of the basic cell values. In the last few years, the prevailing of pre-training and fine-tuning strategy boosts the performance of language models in a variety of tasks \cite{devlin-etal-2019-bert, liu2019roberta}. Following the trend, a good number of previous works employ the same paradigm with Transformer as the primary backbone in tabular data representation \cite{xiang:2020:TURL, yin-etal-2020-tabert, herzig-etal-2020-tapas, iida-etal-2021-tabbie}. TABERT and TAPAS \cite{yin-etal-2020-tabert, herzig-etal-2020-tapas} linearize the table components and jointly pre-train the text-table pairs with masked language model objective to resolve the table QA problem. TURL \cite{xiang:2020:TURL} devises a novel Masked Entity Recovery objectives to learn the the factual knowledge about entities in relational tables. To encode tables of larger size, TABBIE~\cite{iida-etal-2021-tabbie} reduces the input length with two Transformers to encode rows and columns separately. Besides relational tables, TUTA \cite{Wang2021TUTATT} deals with the other generally- structured tables with tree-based attention to extract spatial and hierarchical information. 

The pre-trained tabular models can be utilized in diverse down-stream tasks from Table QA to column type annotation. However, as discussed in \refsec{sec:baseline}, many models don't have the capability on cell level data selection (\eg, \cite{yin-etal-2020-tabert}) and handle numeric values poorly (\eg, \cite{icdm2019}). Even the models that could provide cell embedding vectors for referenced data selection (\eg, \cite{herzig-etal-2020-tapas,2019TabFact}) lack proper cell sampling strategy, because they were designed for table QA and table fact verification tasks, where usually few cells are selected based on the query or the statements. Our framework advances the line of research in flexible representation of tables adaptive to the distinguishing requirements for distinct tasks.

\section{Conclusion}
\label{sec:conclusion}
In this paper, we propose to facilitate intelligent table analysis and visualization by understanding the user motivation in terms of user intent and data focus, denoted as analytical semantics. We design the ASTA framework to apply analytical semantics to automatic recommendation of multiple table analyses and visualizations, \eg, conditional formatting and chart recommendation tasks. The ASTA framework characterize data through statistical and linguistic modules, and allows multi-level data referencing through pre-trained language and tabular models. To evaluate the conditional formatting recommendation, presented for the first time in this study, we collect a large corpus ConFormT with 54k tables and 289k conditional formatting records. Experiments show the good performance of ASTA framework on ConFormT and public chart corpora, validating the effectiveness and generality of our method.


\bibliographystyle{ACM-Reference-Format}
\bibliography{references}

\end{document}